HYPOTHESIS AND THEORY MANUSCRIPT

# Does inter-organellar proteostasis impact yeast quality and performance during beer fermentation?

Bianca de Paula Telini, Marcelo Menoncin and Diego Bonatto*

Brewing Yeast Research Group, Centro de Biotecnologia da UFRGS, Departamento de Biologia Molecular e Biotecnologia, Universidade Federal do Rio Grande do Sul, Porto Alegre, RS, Brazil

**Running title:** Brewing yeast proteostasis

**\*Corresponding author:**
Diego Bonatto
Centro de Biotecnologia da UFRGS - Sala 107
Departamento de Biologia Molecular e Biotecnologia
Universidade Federal do Rio Grande do Sul – UFRGS
Avenida Bento Gonçalves 9500 - Prédio 43421
Caixa Postal 15005
Porto Alegre – Rio Grande do Sul
BRAZIL
91509-900
Phone: (+55 51) 3308-7765
E-mail: diegobonatto@gmail.com

Contract/grant sponsor: CNPq

**Abstract**

During beer production, yeast generate ethanol that is exported to the extracellular environment where it accumulates. Depending on the initial carbohydrate concentration in the wort, the amount of yeast biomass inoculated, the fermentation temperature, and the yeast attenuation capacity, a high concentration of ethanol can be achieved in beer. The increase in ethanol concentration as a consequence of the fermentation of high gravity (HG) or very high gravity (VHG) worts promotes deleterious pleiotropic effects on the yeast cells. Moderate concentrations of ethanol (5% v/v) change the enzymatic kinetics of proteins and affect biological processes, such as the cell cycle and metabolism, impacting the reuse of yeast for subsequent fermentation. However, high concentrations of ethanol (>5% v/v) dramatically alter protein structure, leading to unfolded proteins as well as amorphous protein aggregates. It is noteworthy that the effects of elevated ethanol concentrations generated during beer fermentation resemble those of heat shock stress, with similar responses observed in both situations, such as the activation of proteostasis and protein quality control mechanisms in different cell compartments, including endoplasmic reticulum (ER), mitochondria, and cytosol. Despite the extensive published molecular and biochemical data regarding the roles of proteostasis in different organelles of yeast cells, little is known about how this mechanism impacts beer fermentation and how different proteostasis mechanisms found in ER, mitochondria, and cytosol communicate with each other during ethanol/fermentative stress. Supporting this integrative view, transcriptome data analysis was applied using publicly available information for a lager yeast strain grown under in beer production conditions. The transcriptome data indicated upregulation of genes that encode chaperones, co-chaperones, unfolded protein response elements in ER and mitochondria, ubiquitin ligases, proteasome components, *N*-glycosylation quality control pathway proteins, and components of processing bodies (p-bodies) and stress granules (SGs) during lager beer fermentation. Thus, the main purpose of this hypothesis and theory manuscript is to provide a concise picture of how inter-organellar proteostasis mechanisms are connected with one another and with biological processes that may modulate the viability and/or vitality of yeast populations during HG/VHG beer fermentation and serial repitching.

**Introduction**

During beer production, ethanol generated as a by-product of fermentation is exported to the extracellular environment, where it accumulates. Depending on the initial mono-, di-, and trisaccharide concentrations present in the wort, the amount of yeast cell biomass inoculated, fermentation temperature, and the attenuative capability of yeast strains employed by the brewer, a high concentration of ethanol can be achieved in beer (Puligundla et al., 2011).

At present, the brewing industry is trying to implement the use of very high gravity (VHG) worts (24 °P or approximately 1.101 kg.L$^{-1}$ dissolved solids) to produce beer, which can save energy, time, labor, and capital costs, and improve plant efficiency (Silva et al., 2008; Puligundla et al., 2011). Beer produced from VHG worts contains high quantities of ethanol and other volatiles, which are dissolved in oxygen-free water to produce regular beers with 5% (v/v) ethanol (Stewart, 2010). However, the use of VHG worts imposes challenges for serial repitching due to the osmotic and oxidative stresses that yeast cells experience in the first hours of fermentation, which are followed by ethanol, nutritional, and thermal (cold shock) stresses in the later phases of fermentation and the beginning of cold maturation (Gibson et al., 2007). These stress conditions can lead to yeast slurries that display sluggish fermentation and poor viability, which precludes their use in subsequent fermentations (Huuskonen et al., 2010).

The increase in ethanol concentrations, as a consequence of VHG wort fermentation, can have pleiotropic effects in yeast. Ethanol is a chaotropic substance that affects cell macromolecular structures by reducing hydration (Hallsworth, 1998; Cray et al., 2015). Moderate concentrations of ethanol (around 5% v/v) can alter the enzymatic kinetics of proteins associated with primary metabolism (e.g., glycolysis), and affect different biological processes, such as the cell cycle (Hallsworth, 1998). In comparison, high concentrations of ethanol (>5% v/v) can cause substantial changes in the structure and composition of hydrophobic molecules within the cell (Hallsworth, 1998). Thus, by reducing the water activity in the cell, ethanol promotes a water stress condition (Hallsworth, 1998). In a general sense, the effects of high concentrations of ethanol resemble those observed during heat shock conditions (defined as exposure to temperatures >35 °C), and similar responses are observed in response to both stress situations, such as changes in membrane composition and synthesis of small protective osmolytes (e.g., glycerol and trehalose) (Piper, 1995). Interestingly, it was recently demonstrated by transcriptome analysis using RNA-seq data that ethanol tolerance in different *Saccharomyces cerevisiae* strains also depends on a series of environmental conditions (e.g., the presence or absence of dissolved oxygen), pointing to a strain-by-oxygen-by-alcohol interactions that lead to ethanol tolerance (Sardi et al., 2018).

Protein folding and activity, key features of "proteostasis", are strongly affected by ethanol. In this review, proteostasis mechanisms are defined as all steps required for a protein to exert its function(s), from protein biogenesis to degradation, including all post-translational changes that the protein experiences in between.

In mammalian models, it has been shown that post-translational modifications of proteins, like mannosylation and galactosylation, are substantially changed in ER and Golgi after ethanol shock (Ghosh et al., 1995; Esteban-Pretel et al., 2011). In yeast, there is limited available data on how ethanol affects post-translational modification of proteins, but it is clear that protein structure and activity change in the presence of ethanol (Hallsworth, 1998). It has been observed that ethanol can induce heat shock proteins like Hsp104p, Hsp70p, and Hsp26p, and oxidative stress-response proteins, like Ctt1p, Sod1, and Sod2p under moderate concentrations of ethanol (6% v/v) (Stanley et al., 2010). DNA microarray data supports the idea of a fermentative stress response associated with ethanol toxicity in industrial lager fermentations (Gibson et al., 2007).

Despite the paucity of data regarding the effects of ethanol toxicity in the modulation of yeast proteostasis mechanisms during VHG beer fermentation and serial repitching by using publicly available DNA microarray data (Table S1 and Figure S1), we observed the upregulation of genes linked to lager beer fermentation (Figures S2A and B), including differentially expressed genes (DEGs) associated to organellar proteostasis mechanisms in DNA microarray single analysis (Figures S3A and B to S4A and B) and DNA microarray meta-analysis (Figures S5A and B to S6A and B). The Pan-DEGs resulting from both DNA microarray analyses (Figures S1 and S9A and B) include genes linked to ER-associated unfolded protein response (UPR), endoplasmic reticulum-associated protein degradation (ERAD) responses (Figures 1A and B), and mitochondria-associated proteostasis (Figures 2A and B), suggesting cellular cross-talk among organellar proteostasis mechanisms. It is important to note that all three gene expression datasets (GSEs) analyzed in this work (Table S1) employed the Affymetrix Yeast Genome 2.0 Array for transcript detection of both *S. cerevisiae* and *Schizosaccharomyces pombe* yeast species (https://www.ncbi.nlm.nih.gov/geo/query/acc.cgi?acc=GPL2529), which can potentially introduce a bias when transcripts of *Saccharomyces pastorianus* are evaluated due to the hybrid genome of this species (Okuno et al., 2016). In order to evaluate if the parental genomes of *S. pastorianus* display some specific expression pattern, Horinouchi et al. (2010) designated a custom DNA microarray platform for *S. pastorianus* transcriptome analysis containing probes for both *S. cerevisiae* and *Saccharomyces bayanus* genomes. This custom DNA microarray was employed to evaluate gene expression pattern in the lager brewing strain Weihenstephan 34/7 during a pilot-scale fermentation condition. The transcriptome data gathered by the authors indicated a strong correlation between the expression levels of *S. cerevisiae* and *S. bayanus* orthologous genes during fermentation, allowing discriminate only a small set of *S. cerevisiae* or S. *bayanus* DEGs (Horinouchi et al., 2010). On the other hand, the use of RNA sequencing technologies for evaluation of gene expression in *S. pastorianus* strains during beer production is virtually absent, making it difficult to understand the contribution of parental genomes of *S. pastorianus* in ethanol tolerance and proteostasis. Thus, considering the importance of *S. pastorianus* for brewing industry in general and for hybrid yeast species research (Gorter de Vries et al., 2019; Gorter de Vries et al., 2019), it is imperative to design new experimental procedures for the analysis of the influence of hybrid genomes in proteostasis and ethanol tolerance.

**Yeast ER proteostasis and ethanol tolerance**

The endoplasmic reticulum (ER) consists of an extensive network of membranes that originates at the nuclear envelope and flows through the cytoplasm (English and Voeltz, 2013). It is the site of secretory, membrane, lysosomal, and vacuolar protein synthesis. Besides proteins, the ER is also fundamental for the synthesis of lipids and the assembly of lipid bilayers (van Meer et al., 2008). In the ER, proteins are structurally modified, which involves cleavage of signal sequences, *N*-linked glycosylation, disulfide bond formation, folding of monomers and oligomerization (Braakman and Hebert, 2013). Correct protein folding is facilitated by different molecular chaperones and folding enzymes present in the ER, such as protein disulfide isomerases (PDIs). When a protein is unable to fold correctly, an ER quality control (ERQC) system is activated, comprised of both UPR and ERAD mechanisms (Brodsky and Wojcikiewicz, 2009).

Considering that many proteins found in the ER contain *N*-linked glycans, it is logical to consider that proteostasis mechanisms are largely associated with *N*-glycan synthesis in the ER. In fact, *N*-glycan modification by different glycanases found in ER defines the final destination of polypeptides, and the trimming of glucose residues recruit lectin chaperones that facilitate protein folding (Molinari, 2007; Ferris et al., 2014). Until now, data regarding *N*-glycan processing in yeast during VHG beer fermentation or yeast reuse has been extremely limited. However, our transcriptome data single- and meta-analysis (Figure S1) of the proprietary lager yeast CB11 strain (Coors Brewing Limited (Burton on Trent, UK) (Lawrence et al., 2012) under fermentation conditions, when compared to propagation conditions, point to upregulation of genes related to *N*-glycan processing, like *PDI1* and *PMT1*, which are also important components of the ERAD response (Figures 1A to B). ERAD components export unfolded proteins to the cytosol, which are ubiquitinated and degraded by the 26S proteasome (Brodsky and Wojcikiewicz, 2009; Hetz et al., 2015). The recognition step of unfolded protein can occur either on the luminal side (ERAD-L), the cytosolic side (ERAD-C), or inside of the ER membrane (ERAD-M) (Thibault and Ng, 2012). Protein disulfide isomerase 1, or Pdi1p, is essential for cell viability and is highly abundant in the ER (Mizunaga et al., 1990; Pfeiffer et al., 2016). Pdi1p is also involved in the removal of aberrant disulfide bridges (Gilbert, 1997; Pfeiffer et al., 2016). Interestingly, Pdi1p has chaperone activity, even with proteins that do not form disulfide bridges (Pfeiffer et al., 2016), assisting in the unfolding and the export of ERAD-client proteins from the ER (Weissman and Kimt, 1993). Finally, Pmt1p is an *O*-mannosyltransferase that, together with Pmt2p, exerts proteostasis control of ER proteins. Pmt1p interacts with Pdi1p in order to promote the correct folding of ER-resident proteins or to target misfolded proteins to Hrd1p, a major ERAD-associated E3 ubiquitin-protein ligase (Goder and Melero, 2011). It is worth noting that Pdi1p interacts with Htm1p/Mnl1p, an alpha-1,2-specific exomannosidase that generates Man7GlcNac2, an oligosaccharide structure on glycoproteins target for ERAD (Clerc et al., 2009). Moreover, Htm1p/Mnl1p is required for Yos9p activity (Clerc et al., 2009), a 75 kDa soluble ER glycoprotein (Friedmann et al., 2002) that has been shown to have an important role in glycoprotein degradation (Szathmary et al., 2005). It should be point that *YOS9* gene was found overexpressed in DNA microarray single analysis only (Figures S4 and B). The roles of Htm1p/Mnl1p in yeast cells subjected to VHG beer fermentation and/or ethanol stress are poorly understood, but it has been demonstrated that ethanol can impair the biosynthesis of *N*-glycans in liver cell models *in vitro* (Welti and Hülsmeier, 2014). This indicates that *N*-glycan biosynthesis and processing may be negatively affected by ethanol/fermentation stress during VHG or even high gravity (HG) beer production.

In addition to *N*-glycan structural alterations promoted by ethanol, the presence of unfolded proteins in ER reduces or even stops the translation of new proteins, and also exposes sticky hydrophobic amino acids in unfolded proteins, promoting so-called proteotoxicity (Ron, 2002; Mori, 2015), which is sensed by the transmembrane protein Ire1. Ire1p undergoes oligomerization and autophosphorylation and activates the endoribonuclease domain on the cytosolic side of the membrane that removes a regulatory intron in the *HAC1* mRNA (Chapman and Walter, 1997; Sidrauski and Walter, 1997), leading to the translation of active Hac1p, a bZip transcription factor associated with ER proteostasis (Liu and Chang, 2008). Noteworthy, Navarro-Tapia et al. (2017) showed that low to high concentrations of ethanol ($\leq$ 8% v/v) did not promote protein unfolding in yeast cells, but did trigger UPR through an unknown mechanism in laboratory yeast strains cultured in synthetic medium. However, Miyagawa et al. (2014) showed that an increase in ethanol concentration, from 8 to 16% (v/v) in a synthetic culture medium, promoted the constant expression of *HAC1* spliced form mRNA, which demonstrated that UPR can become chronically activated. In addition, the same authors verified that Kar2p associated with unfolded protein aggregates in the ER when yeast cells were challenged with ethanol at a concentration of 16% (v/v), supporting the idea that very high concentrations of ethanol potentially induce protein aggregates in the ER and trigger ERQC (Miyagawa et al., 2014).

Our transcriptome data indicated that Pan-DEGs related to the classical UPR pathway, like *KAR2, PTC2,* and *YPT1,* are upregulated in lager beer fermentation compared to the yeast propagation step (Figures 1A and B). Ptc2p is a type 2C serine/threonine phosphatase that downregulates the UPR mechanism by dephosphorylating Ire2p (Welihinda et al., 1998), while Ypt1p is a yeast Rab1 homolog that interacts with unspliced *HAC1* mRNA and regulates the UPR by promoting the decay of *HAC1* mRNA (Tsvetanova et al., 2012). Ypt1p has been linked to the maintenance of Golgi morphology and protein composition, participates in ER to Golgi anterograde/retrograde transport, and is necessary for intra Golgi transport (Kamena et al., 2008). While anterograde/retrograde ER to Golgi responses have been extensively studied in yeast and other model organisms, and the functions of a number of different protein complexes involved in these processes have been discerned (Lee et al., 2004), the influence of anterograde/retrograde ER to Golgi transport in brewing yeast vitality or beer fermentation is unknown. However, we hypothesize that this mechanism may be negatively modulated by high ethanol concentrations during VHG beer fermentation or yeast reuse. In support of this idea, it was previously shown that the rat PC12 cell line, when subjected *in vitro* to a low alcohol concentration (30 mM), exhibited delayed anterograde ER to Golgi transport, fragmented Golgi morphology, and a decreased number of secretory vesicles (Tomás et al., 2012). Interestingly, 5% of all eukaryotic proteins (referred to as tail-anchored (TA) proteins) possess a unique carboxy-terminal transmembrane region that targets them to the ER membrane (Stefanovic and Hegde, 2007). Considering that these proteins contain a hydrophobic domain that makes them prone to aggregation in the aqueous environment of the ER lumen, they should be targeted to the ER membrane to avoid the formation of protein aggregates. Thus, in order to guide the entry of TA proteins into the ER membrane, the guided entry of TA proteins (GET) pathway mediates the process, also acting in vesicle fusion and retrograde Golgi to ER responses (Denic et al., 2013). Moreover, the GET pathway is necessary for the retrieval of the Erd2p HDEL receptor from the Golgi to the ER (Schuldiner et al., 2005). Erd2p is an important component that retain proteins bearing a C-terminal tetrapeptide HDEL sequence in the ER, like the ER chaperone Kar2p (Semenza et al., 1990), invertase, and many other secreted proteins. In our transcriptome data analysis, we found that during lager beer fermentation, the Pan-DEGs *GET1*, *GET2*, *GET3*, *GET4*, and *SGT2* are significantly upregulated (Figures 1A and B). GET proteins are core components of GET pathway that promote the transfer of TA proteins from ribosomes to the Get4p/Get5p/Sgt2p complex and to the chaperone Get3p (Chartron et al., 2012). Then, Get1p and Get2p, which comprise a transmembrane complex, drive a conformational change that enables the release of TA proteins from Get3p and, as a consequence, insertion into the ER membrane (Wang et al., 2014). In the context of beer fermentation and ethanol stress, we speculate that ethanol generated during fermentation induces conformational changes in *N*-glycans and secreted proteins that potentially leads to the formation of aggregates in the ER, followed by modification of the structure and function of Golgi. This may result in the activation of ERQC mechanisms and promote the retrograde response of Golgi to ER by stimulating the function of the GET pathway (Figure 4). Finally, the induction of ERQC due to ethanol generated during beer fermentation may also occur in cytoplasm and mitochondria, especially due to the activity of multi-organellar ubiquitin ligases and chaperones.

**Cytosol proteostasis in brewing yeast and the impact on beer fermentation**

In the cytosol, misfolded proteins that have exposed hydrophobic amino acid residues are recognized by protein quality control mechanisms (Buchberger et al., 2010). The cytoplasmic proteostasis mechanism in yeast comprises the heat shock response (HSR) (Mager and Ferreira, 1993), which promotes the expression of molecular chaperones and the proteasome system (Parsell et al., 1993). Similar to UPR, the HSR is induced by different stress conditions that lead to proteotoxicity. In *S. cerevisiae*, the HSR is regulated by the heat shock factor 1 (Hsf1p) transcription factor, encoded by the *HSF1* gene (Weindling and Bar-Nun, 2015). Hsf1p promotes an adaptive response to different stressor agents, including ethanol (Weindling and Bar-Nun, 2015).

Yeast cells treated with 6% (v/v) ethanol show induction of Hsf1p activity (Lee et al., 2000), while Hsf1p mutants were defective in ethanol stress-induced target gene expression (Takemori et al., 2006). Interestingly, the ER oxidoreductin, which is encoded by *ERO1* and induces protein disulfide bonds, was upregulated by Hsf1p in yeast cells exposed to ethanol (Takemori et al., 2006), pointing to a crosstalk between HSR and ERQC mechanisms. Unfortunately, the activity of HSR and ERQC in conditions of VHG beer fermentation or yeast serial repitching is not well understood, but we speculate that modulation of the crosstalk between HSR and ERQC mechanisms may promote ethanol tolerance and cell adaptability during beer fermentation. In line with this hypothesis, ubiquitin ligases, which function by transferring ubiquitin to misfolded/unfolded proteins thus targeting them to the 26S proteasome complex, are key components that regulate both HSR and ERQC (Szoradi et al., 2018). It is well known that different organelles have their own specific ubiquitin ligases, such as Hrd1p and Doa10p in the ER (Ruggiano et al., 2014), San1p in the nucleus (Gardner et al., 2005), and Ubr1p, Ubr2p, Hul5p, and Rsp5p in the cytosol (Prasad et al., 2018). However, different ubiquitin ligases have overlapping functions, such as Doa10p in nucleus and cytosol, San1p in cytoplasm, and Ubr1p in the ER (Szoradi et al., 2018). This ubiquitin ligase network is an essential component of inter-organellar proteostasis, yet very little is known about how this communication is mediated. For example, the overexpression of cytosolic Rsp5p, a NEDD4 family E3 ubiquitin ligase, improve thermoresistance and stress tolerance in yeast strains used for bioethanol production (Hiraishi et al., 2006; Shahsavarani et al., 2012). Disruption of *RSP5* increase the production of isoamyl alcohol and isoamyl acetate in laboratory yeast strains (Abe and Horikoshi, 2005). Rsp5p is part of the so-called "Rsp5-ART ubiquitin ligase adaptor network", which acts to promote the endocytosis and degradation of misfolded integral membrane proteins found in the ER, Golgi, and plasma membrane (Zhao et al., 2013). Additionally, Rsp5p interacts with another important cytosolic E3 ubiquitin ligase named Ubr1p, which is a component of the stress-induced homeostatically-regulated protein degradation (SHRED) pathway (Szoradi et al., 2018).

The SHRED pathway is initially activated by transcription of the hydrophilin-coding gene *ROQ1* by different stress conditions due to the presence of Msn2p/4p and Hsf1p-associated stress response elements in the *ROQ1* promoter (Yamamoto et al., 2005; Verghese et al., 2012; Szoradi et al., 2018). Once translated, Roq1p is cleaved by the endopeptidase Ynm3p, and cleaved Roq1p binds to Ubr1p changing its substrate specificity and promoting the degradation of misfolded proteins at the ER membrane and in the cytosol by the proteasome (Szoradi et al., 2018). Ubr1p interacts with the chaperone Hsp70p and with Sse1p, the ATPase component of the heat shock protein Hsp90 chaperone complex (Nillegoda et al., 2010). Moreover, it has been demonstrated that Ubr1p is a fundamental component of ERAD when yeast cells are exposed to heat or ethanol stress, bypassing the functions of the canonical Hrd1p/Der3p and Doa10p (Stolz et al., 2013). Thus, considering the importance of Rsp5p and Ubr1p in heat and ethanol stress response, we hypothesize that under conditions of VHG/HG beer fermentation, the Rsp5-ART ubiquitin ligase adaptor network and SHRED pathway actively target protein aggregates present in the ER and cytosol to ERAD (Figure 4).

Besides ubiquitin ligases, many chaperones are essential to repair and/or prevent misfolded proteins even before they can be targeted to ERAD. In yeast, chaperones are classified in eight distinct families, which are the small heat-shock proteins (SMALL), the AAA+ family, the CCT/TRiC complex, the prefoldin/GimC (PFD) complex, Hsp40, Hsp60, Hsp70, and Hsp90 families (Gong et al., 2009). From transcriptome data analysis, we observed the upregulation of 54 Pan-DEGs linked to chaperone activity (Figures 2A and B) in the lager yeast strain during beer fermentationas compared to the propagation step. Of these 54 Pan-DEGs linked to chaperone activity, 21 Pan-DEGs encode for chaperone proteins that are found in the cytoplasm and mitochondria (Figure 3A) and belong to the HSP70, HSP40, SMALL, AAA+, HSP60, and HSP90 families (Figure 3B).

Considering the chaperones found in cytoplasm that belong to the Hsp70 family, we found that the Pan-DEGs *SSA1-4, SSZ1,* and *SSB2* were upregulated during beer fermentation in comparison to propagation (Figures 2A and B). The roles of Hsp70s proteins in yeast subjected to ethanol stress are extensively documented, including in beer production. For example, it was reported that *FES1*, *SSA2*, *SSA3*, *SSA4*, and *SSE1* are upregulated in a synthetic wort that mimicked a VHG beer fermentation (Qing et al., 2012). Other studies based on proteomics and quantitative RT-qPCR also confirmed the expression of cytosolic Hsp70p during the early phases of beer fermentation in different lager strains (Brejning et al., 2005; Smart, 2007), and it was clearly demonstrated that moderate concentrations of ethanol (>4% v/v) induce the expression of Hsp70 proteins (Piper et al., 1994). In fact, proteins of the Hsp70 family display important functions not only as chaperones, but also in targeting misfolded proteins for proteasome degradation (Kettern et al., 2010; Kim et al., 2013). In addition, Hsp70 proteins form a bi-chaperone system with Hsp104p, a heat shock protein belonging to the AAA+ family (Zolkiewski et al., 2012), and promote the disaggregation and resolubilization of misfolded proteins (Weibezahn et al., 2005). The transcriptome data also indicated that *HSP104* is upregulated during beer fermentation compared to propagation (Figures 2A and B), supporting our hypothesis that ethanol may promote the formation of misfolded protein aggregates in lager yeast strains during beer fermentation, which likely triggers the activity of Hsp70p and Hsp104p to refold and resolubilize the protein aggregates or target them to the proteasome.

Corroborating the importance of *HSP104* for VHG beer fermentation, Rautio et al. (2007) showed that *HSP104* is induced in the first 10-30 hours of fermentation together with *TPS1*, which encodes trehalose phosphate synthase, a key enzyme involved in trehalose biosynthesis and ethanol stress protection (Alexandre et al., 2001) during beer fermentation. The roles of trehalose as a molecular chaperone in protecting yeast cells against protein aggregation are well understood (Singer and Lindquist, 1998) and a synergistic effect of Hsp104p on trehalose accumulation and degradation has been observed (Iwahashi et al., 1998). However, trehalose and Hsp104p are both required when protein aggregation can be reversible in yeast cells (Sethi et al., 2018). It will be interesting to determine if Hsp104p and trehalose act synergistically in VHG beer protecting yeast cells in the early phases of the fermentation process.

In addition to Hsp70 and AAA+ families, we also observed two additional HSP members with high expression in the cytosol of lager yeast cell during beer fermentation compared to cell propagation, which were the SMALL and Hsp40 proteins (Figure 3B). The SMALL or small heat shock proteins/α-crystallin (sHSP) family is comprised of Hsp26p and Hsp42p in *S. cerevisiae*, two proteins important for preventing unfolded protein aggregation that have overlapping functions in non-stressed and stressed yeast cells (Haslbeck et al., 2004). It was previously demonstrated that Hsp26p co-assembles with misfolded proteins and allows the Hsp104p/Hsp70p/Hsp40p complex to disaggregate them (Cashikar et al., 2005). Interestingly, *HSP26* and other HSP-coding genes were found to be upregulated in yeast strains isolated from sherry wines (Aranda et al., 2002), as well as in lager yeast cells in 16 ºP and 24 ºP wort after 24 hours of fermentation (Odumeru et al., 1992). In addition, it was shown that Hsp26P is a key HSP for ethanol production (Sharma, 2001).

Another interesting target of our transcriptome analysis was the *HSP82* Pan-DEG, which was found to be upregulated in lager yeast during beer fermentation (Figures 2A and B), corroborating the previous data of Gibson et al. (2008). Additionally, in brewing yeast, it has been demonstrated by proteomics and transcriptomics that ethanol stress induces the expression of Hsp82p in wine yeasts (Aranda et al., 2002; Navarro-Tapia et al., 2016) and bioethanol yeast strains (Li et al., 2010). The Hsp82 protein, which belongs to the HSP90 family, is an abundant and essential dimeric ATP-dependent chaperone (Borkovich et al., 1989; Richter et al., 2001). It is required to reactivate proteins damaged by heat without participating in *de novo* folding of most proteins (Nathan et al., 1997). Hsp82 target proteins include steroid hormone receptors and kinases (Mayr et al., 2000). It has been demonstrated that Hsp82p is regulated by several co-chaperones, including Aha1p and Hch1p, both of which activate the ATPase function of Hsp82p (Panaretou et al., 2002) and whose Pan-DEGs were found upregulated in lager yeast during beer fermentation (Figures 2A and B). A third co-chaperone named Cpr6p, a peptidyl-prolyl cis-trans isomerase (cyclophilin) that interacts with Hsp82p, and together with Cpr7p, is required for normal yeast growth (Zuehlke and Johnson, 2012). *CPR6* was found to be upregulated in our transcriptome analysis during beer fermentation (Figures 2A and B), but little is known about its roles during beer fermentation. However, protein-protein interaction data (Figure S10) indicate that Cpr6p interacts with Pbp1p, a component of processing bodies (p-bodies) and stress granules (SGs), which may be induced by severe ethanol stress, heat shock, or glucose deprivation (Kato et al., 2011). Induction of p-bodies and SGs by UPR, which has been observed in mammalian cells (Harding et al., 2000; Anderson and Kedersha, 2008) may also occur in yeast cells. In fact, it would be interesting to determine whether p-bodies/SGs are formed during beer fermentation and if they are associated with proteostasis in cytosol and/or the ER. Cpr6p also interacts with Rpd3p (Figure S10), a conserved histone deacetylase that together with Sin3p and Ume1p comprise the Sin3 complex, a global regulator of transcription that is linked to a series of physiological conditions in yeast and other organisms (Silverstein and Ekwall, 2005), such as ethanol stress (Ma and Liu, 2012). Thus, Cpr6p could be an important co-chaperone that together with Hsp82 may serve as a hub for p-bodies/SGs and epigenetic regulation of genes linked to beer fermentation and proteostasis.

**Mitochondrial proteostasis in brewing yeast**

During beer production, yeast mitochondria exert important functions despite the catabolic repression of nuclear genes encoding mitochondrial proteins linked to respiration (O'Connor-Cox et al., 1996). In fact, mitochondria are not only the primary site of lipid and ergosterol synthesis, but they also provide a series of metabolites originating from central carbon and proline-arginine metabolism (Kitagaki and Takagi, 2014). A large proportion of cellular radical molecules are produced as a result of mitochondrial metabolism, which can strongly affect yeast physiology (Kitagaki and Takagi, 2014). Despite the metabolic and physiological importance of mitochondria, mutations linked to the mitochondrial genome that result in *petite* phenotypes can result in the production of off-flavors (related to synthesis of esters and fusel alcohols) in beer fermentation (Ernandes et al., 1993). Finally, in lager yeasts and possibly in ale strains, mitotype can have a strong influence on temperature tolerance (Baker et al., 2019).

Proteostasis in mitochondria includes different chaperones and proteases, as well as proteins that participate in inter-organellar communication, where defects in mitochondrial proteostasis impacts health and aging (Moehle et al., 2019). Similar to ER, mitochondria have a so-called "mitochondrial unfolded protein response" or mtUPR, which was initially characterized in mammalian cells (Zhao, 2002).

Considering that mitochondria have distinct subcompartments within the organelle (e.g., matrix, outer membrane, and intermembrane space), protein import and sorting processes are very complex (Neupert and Herrmann, 2007). Most mitochondrial proteins are imported as unfolded precursors by means of the translocase of outer membrane (TOM) and translocase of inner membrane (TIM) complexes. Upon translocation into the mitochondria, the proteins undergo chaperone-assisted folding (Neupert and Herrmann, 2007).

The transcriptome analysis of lager yeast cells during beer fermentation revealed that TIM-related Pan-DEGs including *TIM8*, *TIM9*, *TIM12*, *TIM17*, and *TIM54* are upregulated (Figures 2A and B). Tim8p and Tim9p belong to the mitochondrial intermembrane space protein transporter complex, which together with Tim10p, Tim12p, and Tim13p, mediates the transit of proteins destined for the inner membrane across the mitochondria intermembrane space (Davis et al., 2007). Tim9p/Tim10p and Tim9p/Tim10p/Tim12p interact with Tim22p, comprising a multioligomeric complex with Tim54p, Tim22p, Tim18p, and Sdh3p (Gebert et al., 2011). The Tim22 complex mediates the insertion of large hydrophobic proteins, like carrier proteins with multiple transmembrane segments, as well as Tim23p, Tim17p, and Tim22p into the inner membrane (Mokranjac and Neupert, 2009). Tim17p is a component of the Tim23 complex, which promotes the translocation and insertion of proteins into the inner mitochondrial membrane (Mokranjac and Neupert, 2009). The Tim23 complex is composed of a membrane-embedded part, which forms the import motor. This component is formed by Tim14p (Pam18p), Tim16 (Pam16p), Tim44p, Mge1p, and mitochondrial Hsp70p (Mokranjac and Neupert, 2009). The Pan-DEGs encoding Pam16p and Pam18p were found to be upregulated in our transcriptome analysis (Figures 2A and B). Despite the large amount of data collected so far about the roles of Tim22 and Tim23 complexes in yeast mitochondria, considerably less is known about their roles in yeast fermentation/ethanol stress. However, Short et al. (2012) showed that yeast temperature sensitive mutant strains for *PAM16* have defects in fermentation linked to lipid metabolism. Moreover, an upregulated Pan-DEG in our transcriptome analysis, *MDJ2*, encode a chaperone belonging to the HSP40 family that regulates Hsp70 chaperone activity and interacts with Pam18p (Mokranjac et al., 2005). In addition, the transcriptome analysis of lager yeast cells revealed upregulation of *TOM6* and *TOM7* Pan-DEGs (Figure 2A and B), both encoding small protein components of the TOM complex (Dekker et al., 1998). At present, the roles of Tom6p and Tom7p in yeast fermentation/ethanol stress remain unknown.

Two important Pan-DEGs found to be upregulated in our transcriptome analysis, *PHB1* and *PHB2* (Figures 2A and B), encode the proteins prohibitin 1 (Phb1p) and 2 (Phb2p), which are part of a large chaperone complex that stabilizes protein structures and is involved in the regulation of yeast replicative life span and mtUPR (Coates et al., 1997; Nijtmans, 2000). In the context of aging and replicative life span, the impact of Phb1p/2p expression during VHG/HG beer fermentation and/or yeast reuse is unknown, despite the fact that a mixed aged yeast population is commonly observed in mostly ale/lager fermentations (Smart et al., 2000; Powell et al., 2003). Moreover, yeast *phb1* and *phb2* mutants are defective in mitochondrial segregation from mother cells to daughter cells, resulting in delayed segregation of mitochondria (Piper et al., 2002). Interestingly, loss of the orthologous prohibitin in *Caenorhabditis elegans* affected the morphology of mitochondria, resulting in fragmented and disorganized structures (Sanz et al., 2003), a phenotype previously observed in yeast strains used for sake (Kitagaki and Shimoi, 2007) and cider (Lloyd et al., 1996) after prolonged anaerobiosis under high concentration of ethanol (>10% v/v). In animal cells, mitochondrial fragmentation is a feature of mitochondrial proteostasis that is activated in response to a high number of misfolded proteins, but that is also observed during mitophagy and programmed cell death (Moehle et al., 2019). Similarly, Fis1p, a protein involved in mitochondria and peroxisome maintenance in yeast, is upregulated when cells are subjected to high ethanol concentrations, thereby promoting mitochondrial fragmentation and inhibition of apoptosis (Kitagaki et al., 2007).

Taking into account mitochondria structure, *YME1* and *AFG3* were also found to be upregulated in lager yeast cells (Figures 2A and B). These Pan-DEGs encode the mitochondrial ATP-dependent metallopeptidase (AAA protease) Yme1p and Afg3p, respectively, which are necessary for degradation of unfolded or misfolded proteins associated with the mitochondrial inner membrane (Arlt et al., 1996; Schreiner et al., 2012). Despite the fact that the specific roles of Yme1p and Afg3p in VHG/HG beer fermentation or yeast reuse are currently unknown, data regarding the modulation of mitochondria activity upon ethanol exposure indicates that ethanol increases oxidative stress and induces the formation of mitochondrial permeability transition (MPT). MPT is a protein structure that forms a pore across the inner and outer membranes of mitochondria, leading to the depolarization of membrane potential, uncoupling of oxidative phosphorylation and ATP depletion, rupture of the outer mitochondrial membrane, and apoptosis induction (Pastorino et al., 1999; Hoek et al., 2002). Interestingly, AAA proteases seem to be essential to coordinate many functions within mitochondria, including mitochondrial genome stability, respiratory chain complexes synthesis, and the mitochondrial membrane metabolism (Patron et al., 2018). Moreover, AAA proteases are essential to modulate the activity of the mitochondrial $Ca^{2+}$ uniporter (MCU) complex. Mutations in mammalian mitochondrial AAA proteases induce constitutive MCU activity and deregulated mitochondrial $Ca^{2+}$ influx, leading to cell death (König et al., 2016). This suggests that yeast AAA proteases may have essential roles in maintaining mitochondrial structure and function during beer fermentation and ethanol stress, and dysfunctions in mitochondrial AAA proteases are likely to affect brewing yeast viability and vitality.

Besides the protein complexes linked to mitochondrial structure and function, our transcriptome analysis revealed an additional eight upregulated genes during beer fermentation (Figure 3A) that encode for mitochondrial molecular chaperones. These included three upregulated Pan-DEGs belonging to the HSP40 family (*MDJ1*, *MDJ2*, and *JAC1*), one to the HSP60 family (*HSP60*), one to the HSP70 family (*ECM10*) and one to the AAA+ family (*HSP78*) (Figures 2A and B). The interaction between mitochondrial Hsp40 and Hsp70 proteins has been extensively documented, being involved in the translocation of proteins to the matrix and folding (Liu et al., 2001). The chaperone Hsp60p is a fundamental protein required to assist the folding and import of different target proteins to the mitochondrial matrix (Reading et al., 1989), also being important for the replication of mitochondrial DNA in yeast (Kaufman et al., 2003). Finally, Hsp78p is a chaperone that displays similar functions with those of the mitochondrial Hsp70 system (Schmitt et al., 1995). Biochemical studies have indicated cooperation between the Hsp70 system and Hsp78p, forming a bichaperone Hsp70-Hsp78 system that assists in protein refolding after stress induction (Krzewska et al., 2001). Similar to Hsp60p, available evidence suggests that Hsp78p is required for the maintenance of mitochondrial genome integrity (Schmitt, 1996). Thus, it is clear that proteostasis mechanisms in mitochondria play a central role in the maintenance of both proteins and mitochondria nucleoid structure and function, the latter of which profoundly affects beer fermentation (Smart, 2007) and hybrid brewing yeast strain adaptability to temperature (Baker et al., 2019).

**Discussion**

Different organelles such as ER, cytosol, and mitochondria display a set of molecules/proteins that are essential for proteostasis under environmental conditions that are prone to induce protein misfolding/unfolding and amorphous aggregate formation, both potentially leading to proteotoxicity. One such condition is beer fermentation, where brewing yeast strains require protection from the toxic and pleiotropic effects of ethanol. In order to deal with ethanol and maintain proteostasis during beer fermentation, the major cellular compartments (e.g., mitochondria, ER, and cytosol) must communicate with one another to mount a systemic cell response (Figure 4).

Multiple lines of evidence indicate that organellar proteostasis is a concerted process that is directly connected with different biological processes, such as metabolism and aging (Raimundo and Kriško, 2018). This so-called "inter-organellar/cross-organellar communication/response" or CORE is dependent on a series of signaling-associated and/or protein networks that include HSPs and their target molecules (Raimundo and Kriško, 2018). Interestingly, one hallmark of the CORE is the upregulation of multiple genes and proteins linked to proteostasis, including *PDI1*, *HSP26*, and *HSP90* (Perić et al., 2016). However, we speculate that other protein and small molecule networks, such as those composed of E3 ubiquitin ligases, the SHRED pathway, trehalose biosynthesis, ERAD, and the prohibitin complex, could be essential components of a larger CORE network that is upregulated during beer fermentation (Figure 4). The activation of a CORE network may impact different aspects of fermentative metabolism that are crucial for yeast viability and/or vitality and further use in serial repitching. For example, it was observed in *C. elegans* that mitochondrial proteotoxicity increases fatty acid synthesis and promotes lipid accumulation, a condition associated with mitochondrial-to-cytosolic stress response that is essential for *C. elegans* survival (Kim et al., 2016). Similarly, we observed in our transcriptome analysis an increase in the expression of genes related to lipid biosynthesis in lager yeast (Figures S11 and S12), pointing to a potentially conserved CORE network in eukaryotes. Furthermore, the roles of inter-organellar proteostasis mechanisms in the replicative and chronological life span of yeast cells have been demonstrated previously (Perić et al., 2016; Chadwick et al., 2019), which are very likely to affect brewing. Finally, a number of important questions remain about how the CORE network may modulate other organelles (e.g., nucleus and vacuole) during beer fermentation (Figure 4). As described above, some components of organellar proteostasis influence transcriptional activity in the nucleus. Recently, Andréasson et al. (2019) demonstrated an important connection between mitochondria and nucleus for proteostasis and cell metabolism. However, little is known about epigenetic modulation during proteotoxic stress induced by ethanol. In the same sense, how the CORE network connects with vacuoles is an open question (Figure 4). Noteworthy, it was demonstrated that in conditions of lipid imbalance, unfolded ER proteins can be removed by lipid droplets and targeted to the vacuole for degradation by microlipophagy (Vevea et al., 2015). However, the impact of this mechanism remains to be determined in beer fermentation.

To evaluate the importance of each component of the CORE network for beer fermentation, it is indispensable to get high quality RNA-seq data from different *S. pastorianus* strains in conditions of industrial yeast propagation and beer fermentation. It can be potentially achieved by tagging the major genes of *S. cerevisiae* and *S. eubayanus* linked to the CORE network followed by interspecies hybridization to generate *S. pastorianus* strains by different techniques, like HyPr (Alexander et al., 2016) and testing them in brewery environment. On the other hand, the use of different proteome techniques to evaluate the contribution of CORE components is also welcome as well as the generation of *S. pastorianus* mutant strains for CORE components by uding CRISPR-Cas9 technology (de Vries et al., 2017).

In conclusion, a better understanding of the CORE network in the context of beer fermentation and/or ethanol stress will allow us to improve different aspects of brewing, from ethanol tolerance in VHG/HG fermentation to yeast reuse, potentially allowing us to select yeast strains with high tolerance to ethanol or diminished aging, which will ultimately improve beer yield and quality.

**Author Contributions**

DB contributed to the design, acquisition, analysis, and interpretation of data for the work; DB, BT, and MM contributed to drafting the work and prepared the final work; DB prepared the figures and all authors approved the final manuscript.

**Conflict of Interest Statement**

The authors declare that this work was conducted in the absence of any commercial or financial relationships that could be construed as a potential conflict of interest.

**Figures legends**

**Figure 1.** (A) Differentially upregulated Pan-genes associated with proteostasis observed in the lager yeast CB11 strain during beer fermentation. The mean expression values are indicated by log2 fold change ± standard deviation (SD) on the y-axis and in the inset. Gene names are indicated on the x-axis. (B) Heatmap plot showing the clustered differentially upregulated genes associated with proteostasis observed in CB11 during beer fermentation and the associated clustered biological processes from gene ontology analysis (see Figure S1). Heatmap rows and columns were grouped using the Euclidean distance method and complete linkage.

**Figure 2.** (A) Differentially upregulated Pan-genes associated with chaperones and folding proteins observed in the lager yeast CB11 strain during beer fermentation. The mean expression values are indicated by log2 fold change ± standard deviation (SD) on the y-axis and in the inset. Gene names are indicated on the x-axis. (B) Heatmap plot showing the clustered differentially upregulated genes associated with chaperones and folding proteins observed in CB11 during beer fermentation and the associated clustered biological processes from gene ontology analysis (see Figure S1). Heatmap rows and columns were grouped using the Euclidean distance method and complete linkage.

**Figure 3.** (A) Number of chaperones and folding protein coding Pan-genes found to be upregulated in different organelles of the lager yeast CB11 strain during beer fermentation, in comparison to yeast propagation. (B) Number of coding Pan-genes upregulated in CB11 during beer fermentation, in comparison to yeast propagation, that are linked to the major chaperone protein families.

**Figure 4.** A model for inter-organellar/cross-organellar communication/response proteostasis (CORE network) in brewing yeast. During beer fermentation and/or yeast reuse, the endoplasmic reticulum (ER), mitochondria, and cytosol regulate proteostasis/protein quality by monitoring their environments and communicating with one another by means of the CORE network. In conditions of proteotoxicity induced by ethanol during beer fermentation, the CORE network is activated and is composed of different proteins/pathways, such as heat shock proteins (HSPs), endoplasmic reticulum-associated protein degradation (ERAD), the stress-induced, homeostatically regulated protein degradation (SHRED) pathway, E3 ubiquitin ligases, and the prohibitin complex. Trehalose, a molecular chaperone necessary for proteotoxic response, is also part of the CORE network. Additionally, each organelle has its own particular mechanisms of protein quality control/proteostasis. The impact of the CORE network in the proteostasis response of vacuoles of brewing yeast is not well understood, but may be associated with microlipophagy. Finally, proteotoxicity induced by ethanol regulates transcriptional activity and epigenetic mechanisms in the nucleus, which are influenced by CORE network components. Moreover, the CORE network activity and proteotoxicity are potentially linked to aging in brewing yeast cells.

Figure 1.

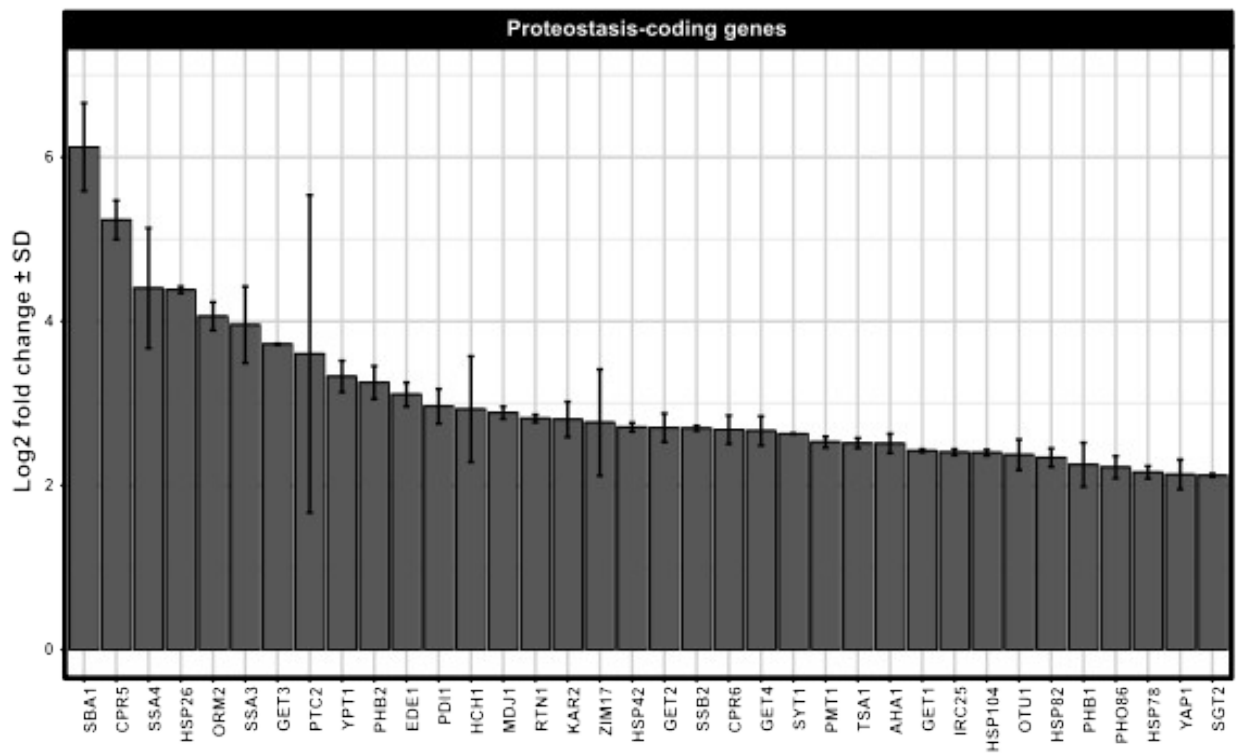

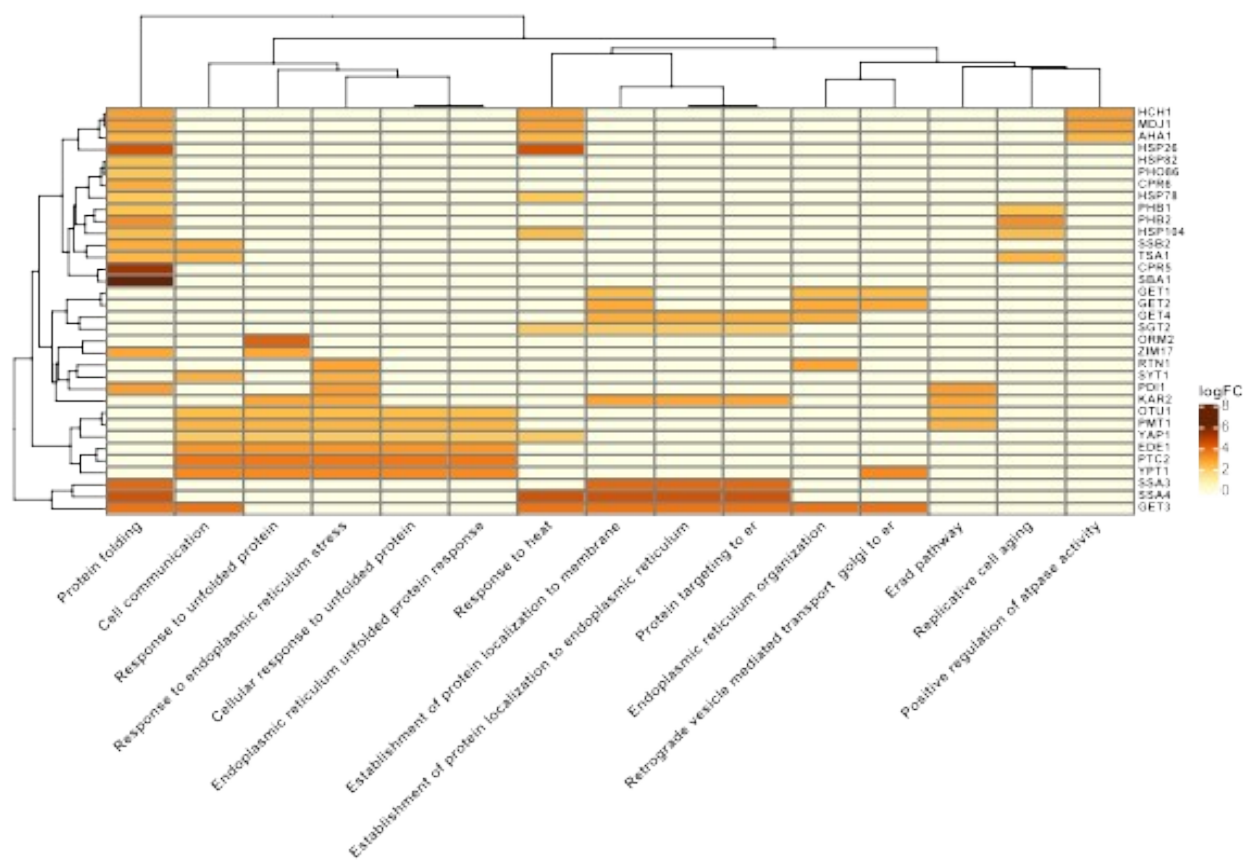

Figure 2.

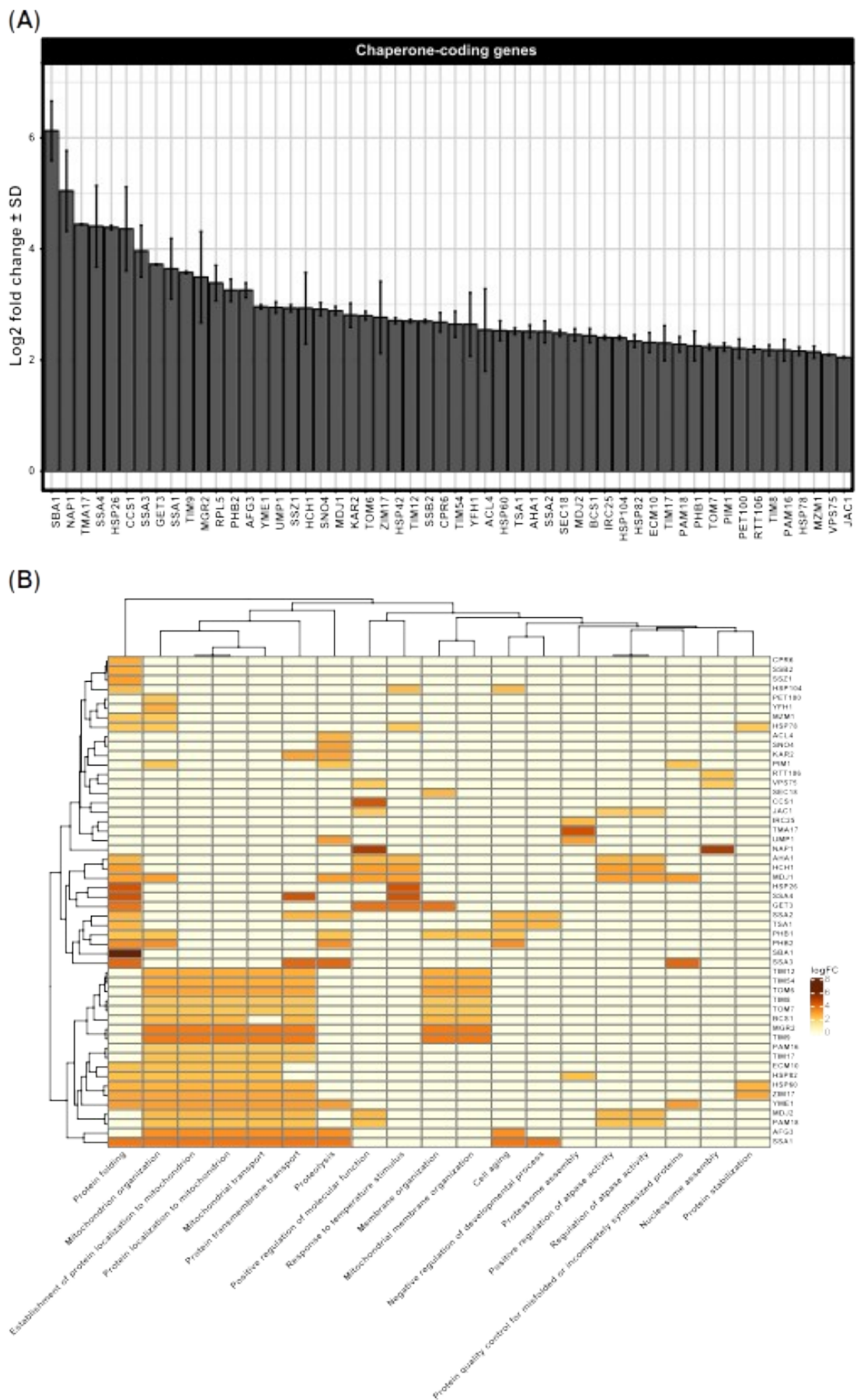

Figure 3.

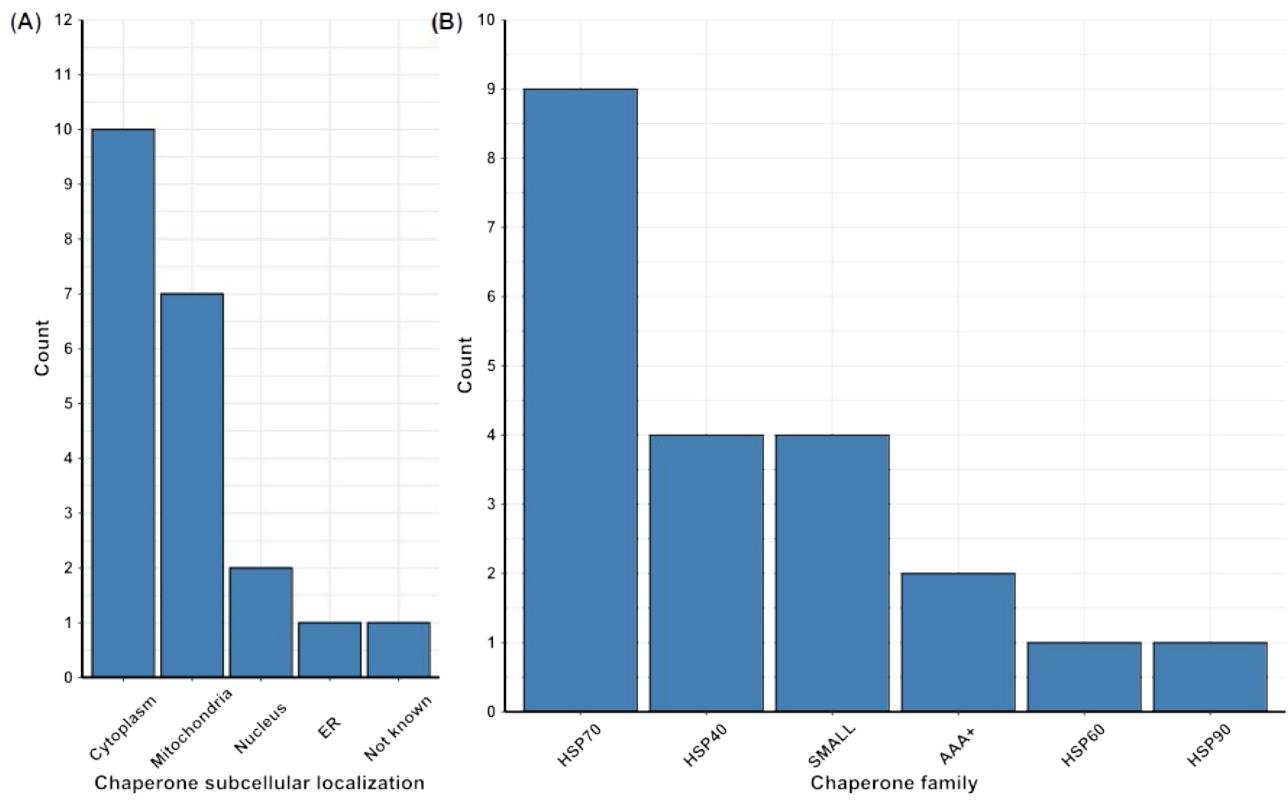

Figure 4.

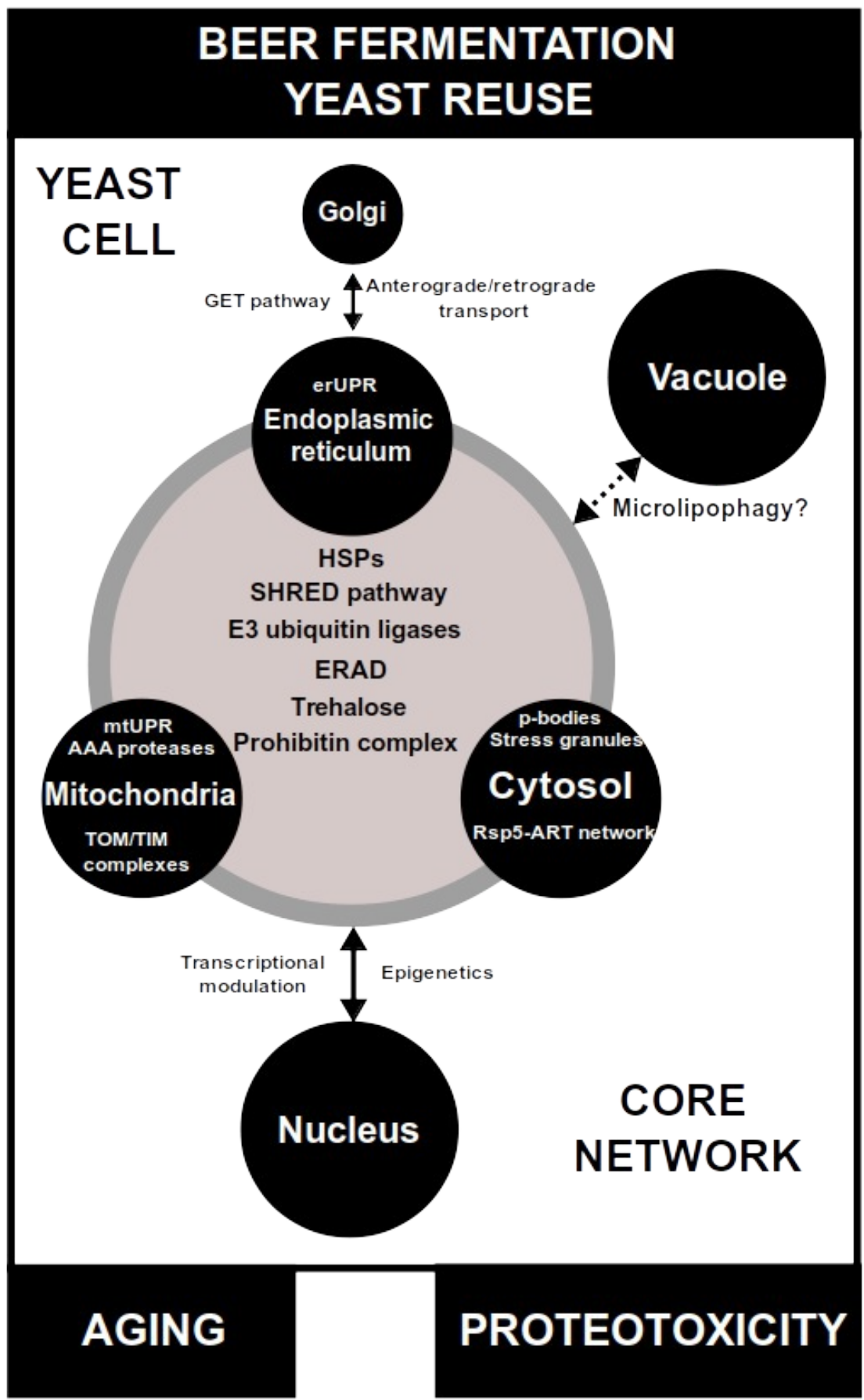

# SUPPLEMENTARY MATERIAL 1

**Experimental procedures**

*DNA microarray gene expression and gene ontology analysis*

DNA microarray gene expression (GSE) datasets (GSE9423, GSE10205, and GSE16376) comparing lager yeast CB11 strain (*Saccharomyces pastorianus*) during fermentation and propagation in different times (Table S1 and Figure S1) (Gibson et al., 2008) were obtained from Gene Expression Omnibus (GEO) database [http://www.ncbi.nlm.nih.gov/gds] (Table S1). In this study, the definitions of fermentation and propagation are the same used in the work by Gibson et al. (2008), where 'fermentation' is the process when beer is made from wort by yeast fermentation in absence of oxygen, while 'propagation' is defined as the process in which a sufficient yeast biomass is generated for beer production using wort as liquid media constantly supplied with molecular oxygen.

**Table S1.** Gene expression datasets (GSEs) used in this work.

| Source[a] | GEO samples files | Sample name | Sample | Organism | Strain |
|---|---|---|---|---|---|
| GSE9423 | GSM239499 | Fermentation 30 hours | Ferm_30h_B | *Saccharomyces pastorianus* | CB11 |
| GSE9423 | GSM239503 | Fermentation 60 hours | Ferm_60h_C | *Saccharomyces pastorianus* | CB11 |
| GSE9423 | GSM239512 | Propagation 30 hours | Prop_30h_C | *Saccharomyces pastorianus* | CB11 |
| GSE9423 | GSM239504 | Fermentation 8 hours | Ferm_8h_A | *Saccharomyces pastorianus* | CB11 |
| GSE9423 | GSM239514 | Propagation 8 hours | Prop_8h_B | *Saccharomyces pastorianus* | CB11 |
| GSE9423 | GSM239501 | Fermentation 60 hours | Ferm_60h_A | *Saccharomyces pastorianus* | CB11 |
| GSE9423 | GSM239510 | Propagation 30 hours | Prop_30h_A | *Saccharomyces pastorianus* | CB11 |
| GSE9423 | GSM239515 | Propagation 8 hours | Prop_8h_C | *Saccharomyces pastorianus* | CB11 |
| GSE9423 | GSM239506 | Fermentation 8 hours | Ferm_8h_C | *Saccharomyces pastorianus* | CB11 |
| GSE9423 | GSM239505 | Fermentation 8 hours | Ferm_8h_B | *Saccharomyces pastorianus* | CB11 |
| GSE9423 | GSM239511 | Propagation 30 hours | Prop_30h_B | *Saccharomyces pastorianus* | CB11 |
| GSE9423 | GSM239502 | Fermentation 60 hours | Ferm_60h_B | *Saccharomyces pastorianus* | CB11 |
| GSE9423 | GSM239508 | Propagation 0 hours | Prop_0h_B | *Saccharomyces pastorianus* | CB11 |
| GSE9423 | GSM239500 | Fermentation 30 hours | Ferm_30h_C | *Saccharomyces pastorianus* | CB11 |
| GSE9423 | GSM239507 | Propagation 0 hours | Prop_0h_A | *Saccharomyces pastorianus* | CB11 |
| GSE9423 | GSM239509 | Propagation 0 hours | Prop_0h_C | *Saccharomyces pastorianus* | CB11 |

| Source[a] | GEO samples files | Sample name | Sample | Organism | Strain |
|---|---|---|---|---|---|
| GSE9423 | GSM239513 | Propagation 8 hours | Prop_8h_A | *Saccharomyces pastorianus* | CB11 |
| GSE10205 | GSM257787 | Fermentation 102 hours | Ferm_102h_A | *Saccharomyces pastorianus* | CB11 |
| GSE10205 | GSM257776 | Fermentation 8 hours | Ferm_8h_A | *Saccharomyces pastorianus* | CB11 |
| GSE10205 | GSM257778 | Fermentation 8 hours | Ferm_8h_C | *Saccharomyces pastorianus* | CB11 |
| GSE10205 | GSM257789 | Fermentation 102 hours | Ferm_102h_C | *Saccharomyces pastorianus* | CB11 |
| GSE10205 | GSM257780 | Fermentation 30 hours | Ferm_30h_B | *Saccharomyces pastorianus* | CB11 |
| GSE10205 | GSM257782 | Fermentation 60 hours | Ferm_60 h_B | *Saccharomyces pastorianus* | CB11 |
| GSE10205 | GSM257786 | Fermentation 80 hours | Ferm_80h_C | *Saccharomyces pastorianus* | CB11 |
| GSE10205 | GSM257785 | Fermentation 80 hours | Ferm_80h_B | *Saccharomyces pastorianus* | CB11 |
| GSE10205 | GSM257781 | Fermentation 60 hours | Ferm_60h_A | *Saccharomyces pastorianus* | CB11 |
| GSE10205 | GSM257783 | Fermentation 60 hours | Ferm_60h_C | *Saccharomyces pastorianus* | CB11 |
| GSE10205 | GSM257784 | Fermentation 80 hours | Ferm_80h_A | *Saccharomyces pastorianus* | CB11 |
| GSE10205 | GSM257777 | Fermentation 8 hours | Ferm_8h_B | *Saccharomyces pastorianus* | CB11 |
| GSE10205 | GSM257779 | Fermentation 30 hours | Ferm_30h_A | *Saccharomyces pastorianus* | CB11 |
| GSE10205 | GSM257788 | Fermentation 102 hours | Ferm_102h_B | *Saccharomyces pastorianus* | CB11 |
| GSE16376 | GSM410831 | Propagation 0 hours | Prop_0h_A | *Saccharomyces pastorianus* | CB11 |
| GSE16376 | GSM410832 | Propagation 0 hours | Prop_0h_B | *Saccharomyces pastorianus* | CB11 |
| GSE16376 | GSM410833 | Propagation 0 hours | Prop_0h_C | *Saccharomyces pastorianus* | CB11 |
| GSE16376 | GSM410834 | Propagation 4 hours | Prop_4h_A | *Saccharomyces pastorianus* | CB11 |
| GSE16376 | GSM410835 | Propagation 4 hours | Prop_4h_B | *Saccharomyces pastorianus* | CB11 |
| GSE16376 | GSM410836 | Propagation 4 hours | Prop_4h_C | *Saccharomyces pastorianus* | CB11 |
| GSE16376 | GSM410837 | Propagation 8 hours | Prop_8h_A | *Saccharomyces pastorianus* | CB11 |
| GSE16376 | GSM410838 | Propagation 8 hours | Prop_8h_B | *Saccharomyces pastorianus* | CB11 |
| GSE16376 | GSM410839 | Propagation 8 hours | Prop_8h_C | *Saccharomyces pastorianus* | CB11 |
| GSE16376 | GSM410840 | Propagation 30 hours | Prop_30h_A | *Saccharomyces pastorianus* | CB11 |
| GSE16376 | GSM410841 | Propagation 30 hours | Prop_30h_B | *Saccharomyces pastorianus* | CB11 |
| GSE16376 | GSM410842 | Propagation 30 hours | Prop_30h_C | *Saccharomyces pastorianus* | CB11 |

[a]In blue color, the GSE9423 used for single DNA microarray analysis. In red, the GSE10205 and GSE16376 used for DNA microarray meta-analysis.

All statistical analyses on transcriptome data were performed using the R platform [https://www.r-project.org] and the following packages: (i) GEOquery for data matrix importing and parsing (Davis and Meltzer, 2007); (ii) arrayQualityMetrics for microarray quality analysis (Kauffmann et al., 2009) and (iii) limma for differentially expressed gene (DEG) analysis (Ritchie et al., 2015) (Figure S1). The significance of DEGs was determined by False Discovery Rate (FDR) algorithm, implemented in limma package (Ritchie et al., 2015). Beer fermentation-associated DEGs from DNA microarray single- (GSE9423) and meta-analysis (GSE10205 versus GSE16376) with mean |logFC| ≥ 2.0, logFC standard deviation (SD) < 1.0 and FDR < 0.05 were selected for gene ontology analyses and protein subcellular localization (Figure S1). To perform the gene ontology analysis, DEGs were specifically filtered for annotated proteostasis- and chaperones-linked genes using data from *Saccharomyces* Genome Database (Figure S1). Further, the proteostasis- and chaperones-linked DEGs obtained from DNA microarray single- and meta-analysis were applied to select a list of commonly observed DEGs in both analysis, and were called as proteostasis Pan-DEGs (Figure S1). The major biological processes and cellular component associated to proteostasis- and chaperones-linked DEGs lists from DNA microarray single- and meta-analysis and Pan-DEGs were further determined using the R package clusterProfile and *Saccharomyces cerevisiae* protein data from UniProt (Yu et al., 2012) (Figure S1). The degree of functional enrichment for a given biological process category was quantitatively assessed ($p$-value < 0.01) using a hypergeometric distribution. Multiple test correction was also assessed by applying FDR algorithm (Benjamini and Hochberg, 1995) at a significance level of $p < 0.05$. Semantic comparison among biological processes and cellular component associated to DEGs were made using R package GOSemSim (Yu et al., 2010) using false discovery rate (FDR) < 0.01 and q-value < 0.05 (Figure S1). Networks containing the subcellular targets of proteostasis- and chaperones-associated DEGs from DNA microarray single analysis, meta-analysis and Pan-DEGs were generated using the R package igraph and Cytoscape 3.7.2 (Shannon et al., 2003; Csardi and Nepusz, 2006, [CSL STYLE ERROR: reference with no printed form.]) (Figure S1). Heatmaps combining proteostasis- and chaperones-associated DEGs from DNA microarray single- and meta-analysis and Pan-DEGs values and GOs were designed with R package ComplexHeatmap (Gu et al., 2016), where rows and columns were grouped using Euclidean distance method and complete linkage (Figure S1). All Figures displayed in this supplementary material as well as in the manuscript can be downloaded at https://github.com/bonattod/Proteostasis_data_analysis.git

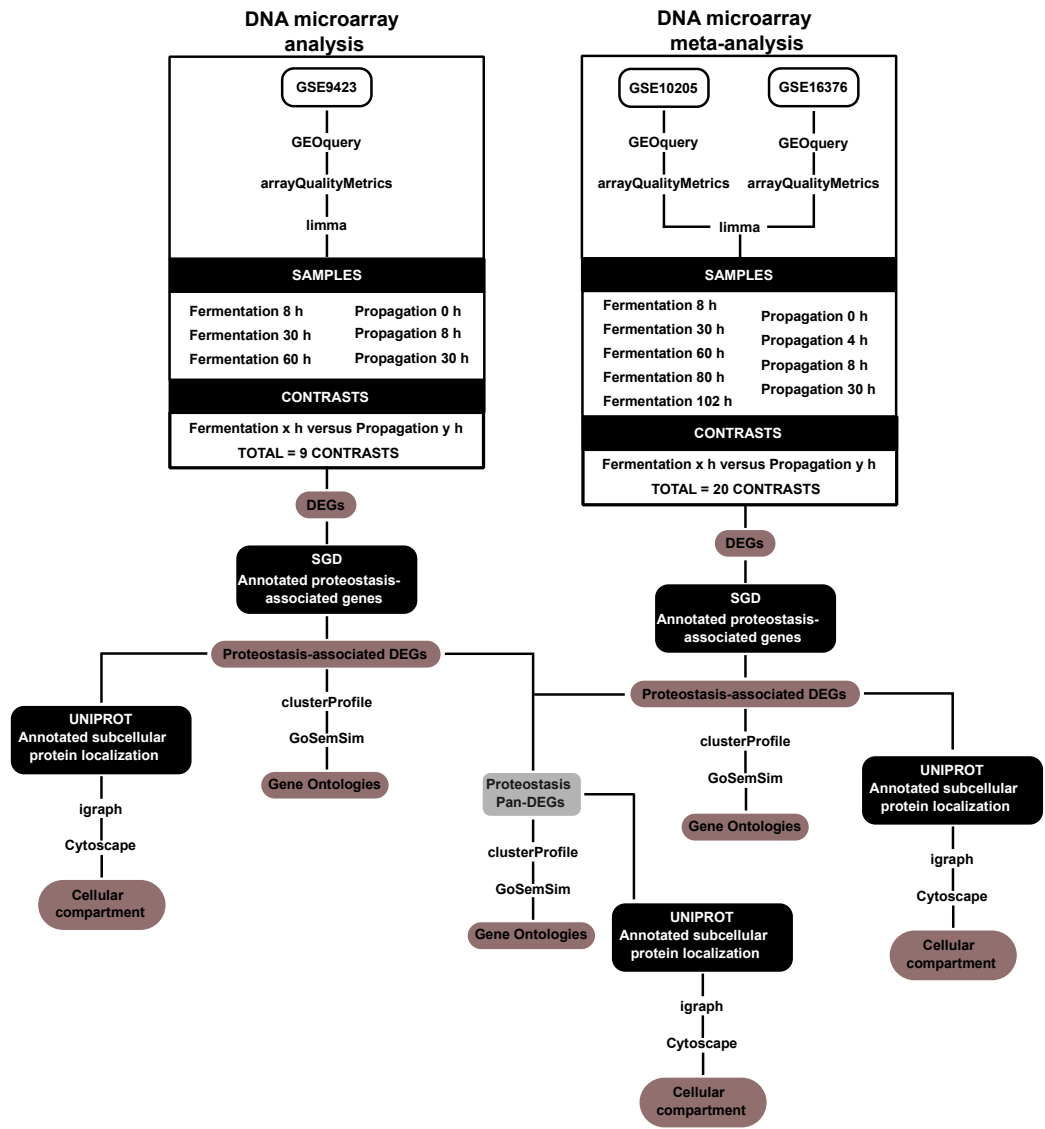

**Figure S1.** Experimental design used in DNA microarray single and meta-analysis. Abbreviation: differential by expressed genes (DEGs); *Saccharomyces* Genome Database (SGD); Universal Protein Resource (UniProt).

**Supplementary results**

*DNA microarray single- and meta-analysis*

Data gathered from DNA microarray single analysis (GSE9423) showed a low number of overexpressed and underexpressed DEGs comparing the very beginning of propagation condition (0 hour) with different times of fermentation (8, 30, and 60 hours) (Figure S2A), while the number of DEGs in fermentation compared to different propagation times (8 and 30 hours) dramatically increased (Figure S2A). A similar result was also observed for DNA microarray meta-analysis comparing different times of fermentation (GSE10205) and propagation (GSE16376) (Figure S2B), where the number of DEGs was low when yeast cells in the first hours of propagation (0 and 4 hours) were compared with yeasts in different times of fermentation (from 8 to 102 hours). Additionally, the number of DEGs in both DNA microarray single- and meta-analysis sharply increased when cells in different times of fermentation were compared with the same yeast strain after 8 hours of propagation (Figures S2A and B).

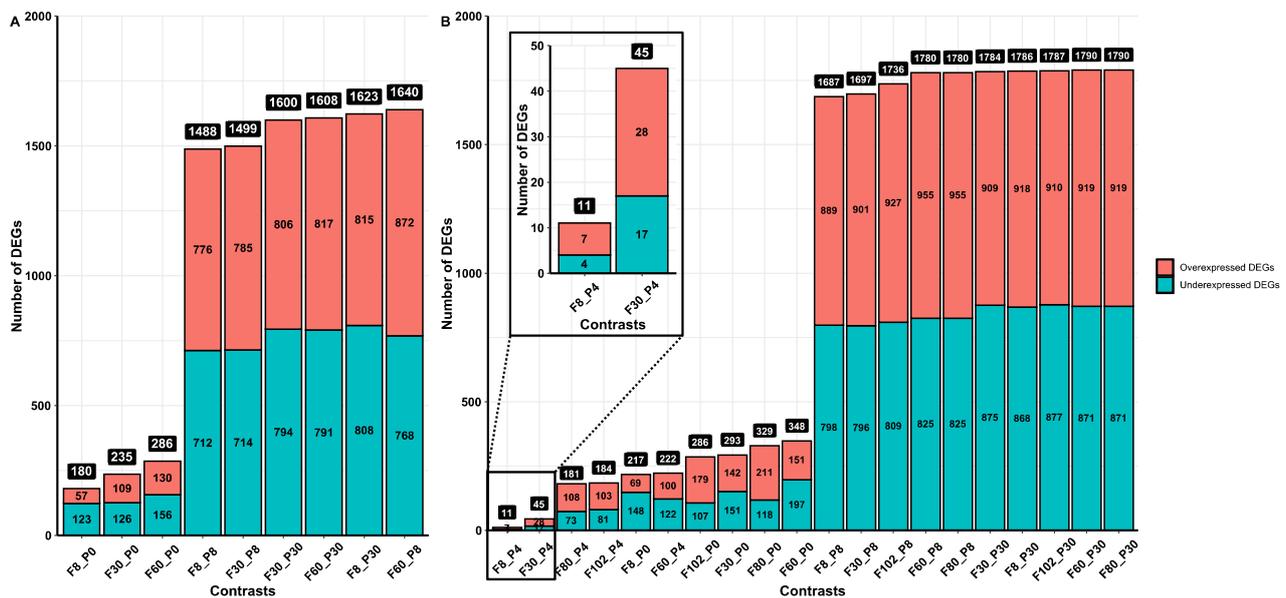

**Figure S2.** In (A), DNA microarray single analysis of GSE9423 dataset comparing the lager yeast CB11 strain in different times of fermentation (F) and propagation (P). In (B), DNA microarray meta-analysis comparing the lager yeast CB11 strain in different times of fermentation (F; GSE10205) and propagation (P; GSE16376). The time of point collection (in hours) is indicated after the letters "F" and "P". The black squares above the bars indicate the total number of DEGs observed for a given contrast. The numbers inside the blue and red bars shown the total of underexpressed and overexpressed DEGs observed in a specific contrast. The inset in the graphic (B) is a zoom of the first two bars.

*Proteostasis- and chaperones-associated DEGs in DNA microarray single- and meta-analysis*

The overexpressed DEGs from DNA microarray single and meta-analysis (Figure S2A and B) were filtered for proteostasis- and chaperones-associated genes using the annotated data from *Saccharomyces* Genome Database (Figure S1). The number of overexpressed proteostasis- and chaperones-associated DEGs observed in beer fermentation using DNA microarray single analysis (Figures S3A and S4A) and DNA microarray meta-analysis (Figure S5A and S6A) was similar. These DEGs were then subjected to a gene ontology (GO) analysis and the major biological processes were evaluated (Figures S3B to S6B). Data from GO analysis showed that similar biological processes were obtained after semantic reduction for both proteostasis- and chaperones-associated DEGs gathered from different DNA microarray analysis (Figures S3B to S6B).

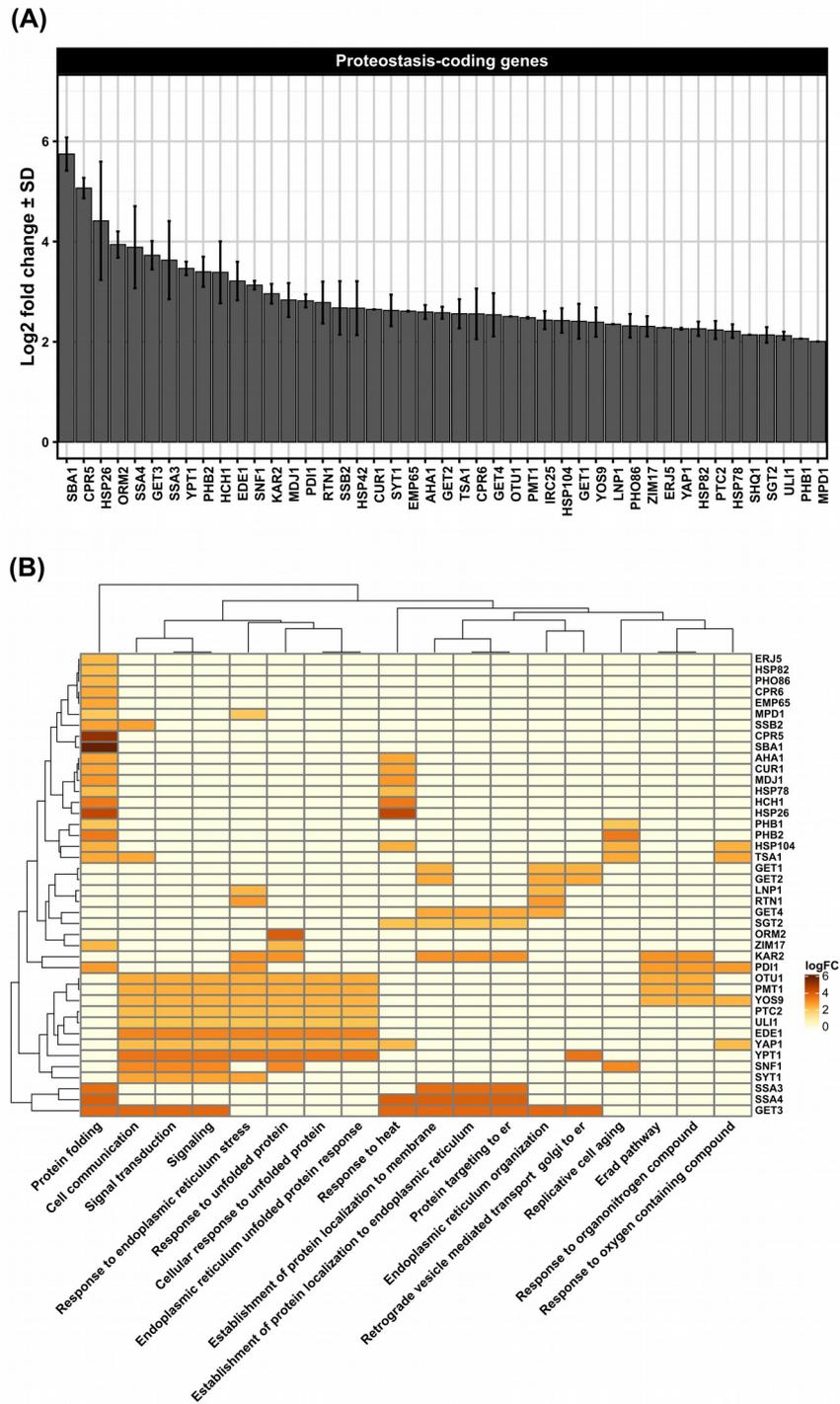

**Figure S3.** (A) Differentially upregulated genes from DNA microarray single analysis (GSE9423) associated with proteostasis observed in the lager yeast CB11 strain during beer fermentation, compared to the propagation step, at different times. The mean expression values are indicated by log2 fold change ± standard deviation (SD) on the y-axis and in the inset. Gene names are indicated on the x-axis. (B) Heatmap plot showing the clustered differentially upregulated genes associated with proteostasis observed in CB11 during beer fermentation, compared to the propagation step, at different times and the associated clustered biological processes from gene

ontology analysis. Heatmap rows and columns were grouped using the Euclidean distance method and complete linkage.

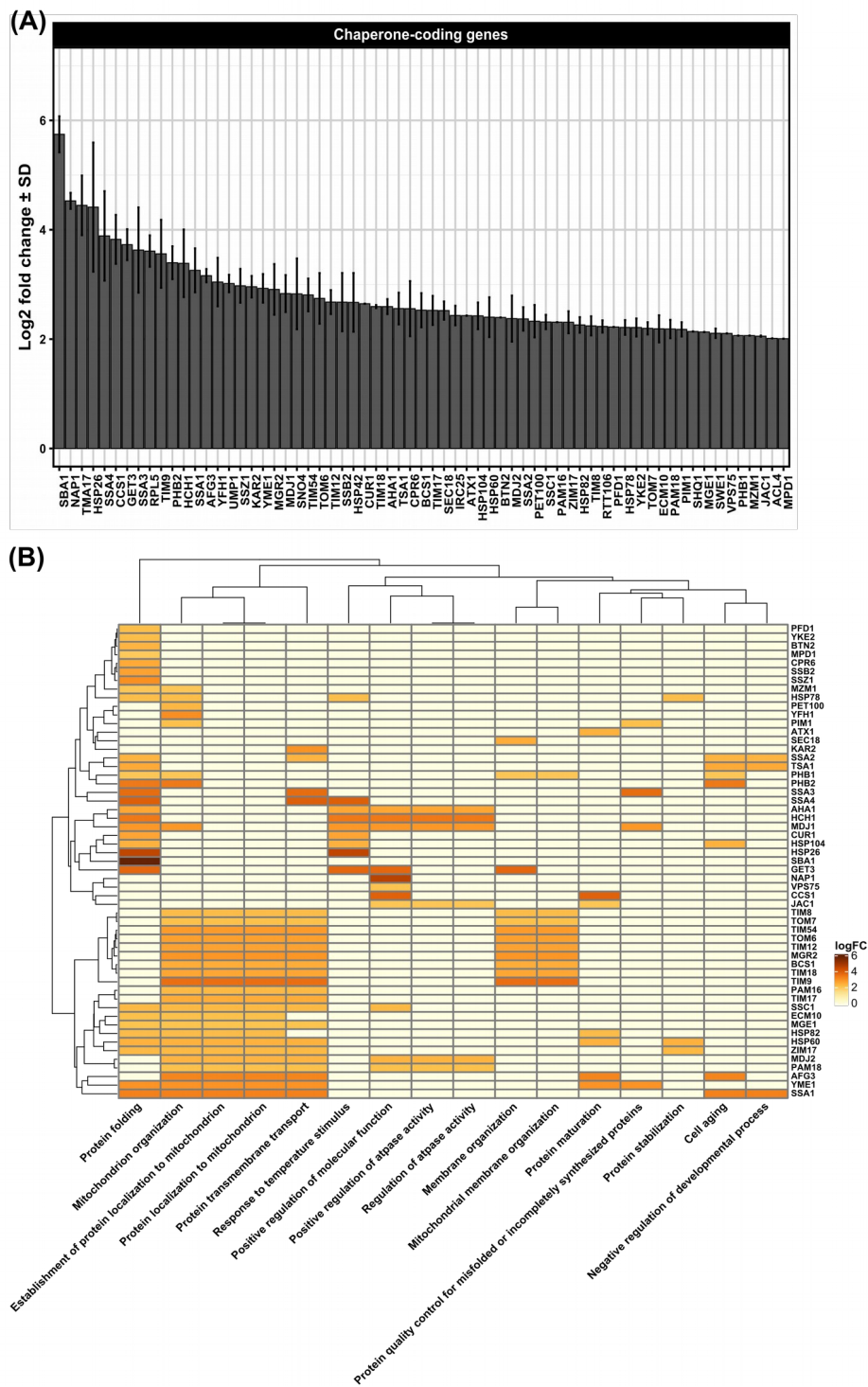

**Figure S4.** (A) Differentially upregulated genes from DNA microarray single analysis (GSE9423) associated with chaperones and folding proteins observed in the lager yeast CB11 strain during beer fermentation, compared to the propagation step, at different times. The mean expression values are indicated by log2 fold change ± standard deviation (SD) on the y-axis and in the inset. Gene names are indicated on the x-axis. (B) Heatmap plot showing the clustered differentially

upregulated genes associated with chaperones and folding proteins observed in CB11 during beer fermentation, compared to the propagation step, at different times and the associated clustered biological processes from gene ontology analysis. Heatmap rows and columns were grouped using the Euclidean distance method and complete linkage.

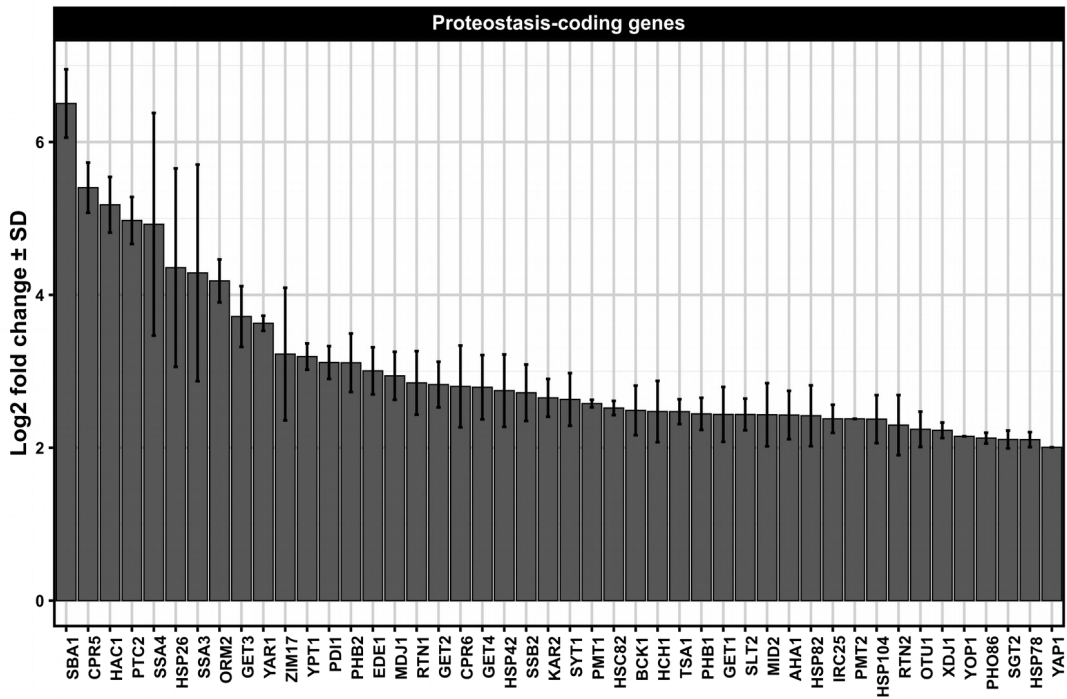

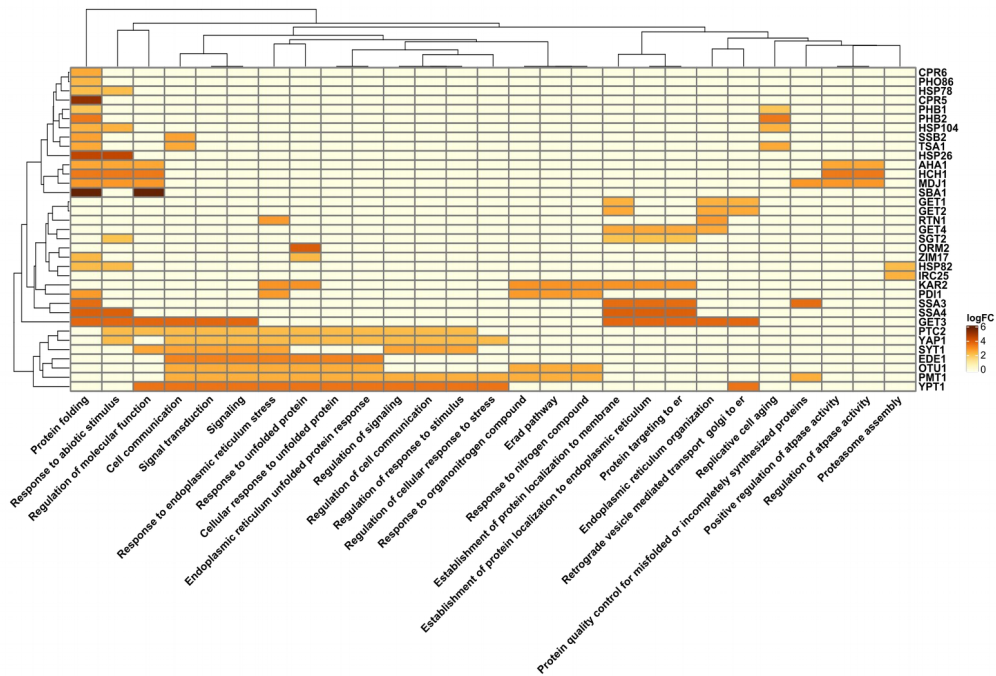

**Figure S5.** A) Differentially upregulated genes from DNA microarray meta-analysis (GSE10205 versus GSE16376) associated with proteostasis observed in the lager yeast CB11 strain during beer fermentation, compared to the propagation step, at different times. The mean expression values are indicated by log2 fold change ± standard deviation (SD) on the y-axis and in the inset. Gene names are indicated on the x-axis. (B) Heatmap plot showing the clustered differentially upregulated genes associated with proteostasis observed in CB11 during beer fermentation, compared to the propagation step, at different times and the associated clustered biological processes from gene ontology analysis. Heatmap rows and columns were grouped using the Euclidean distance method and complete linkage.

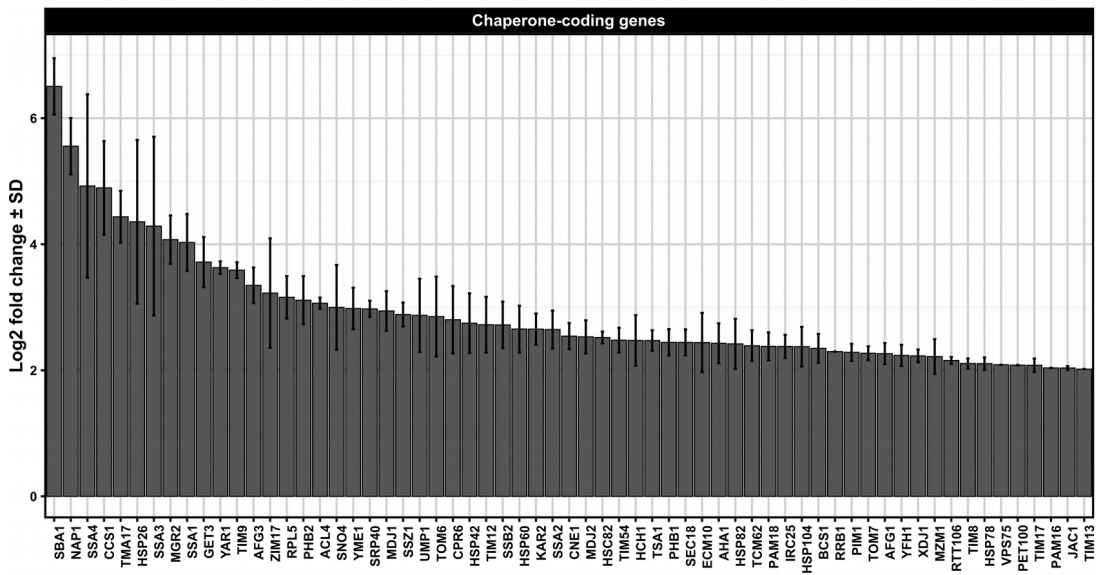

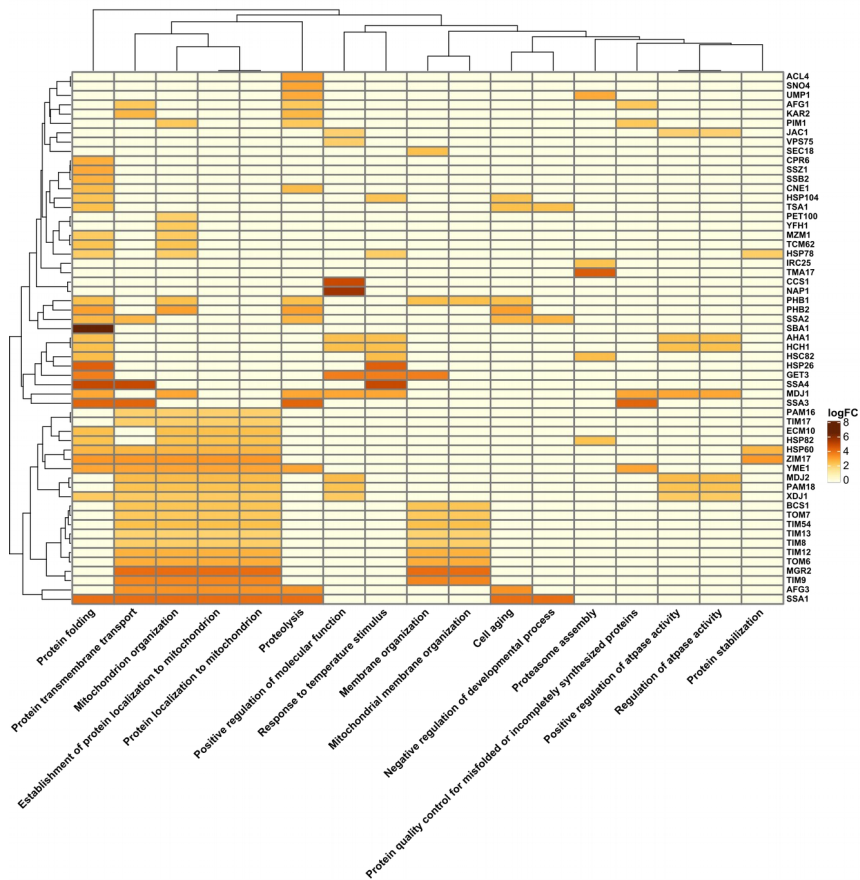

**Figure S6.** (A) Differentially upregulated genes from from DNA microarray meta-analysis (GSE10205 versus GSE16376) associated with chaperones and folding proteins observed in the lager yeast CB11 strain during beer fermentation, compared to the propagation step, at different times. The mean expression values are indicated by log2 fold change ± standard deviation (SD) on the y-axis and in the inset. Gene names are indicated on the x-axis. (B) Heatmap plot showing the clustered differentially upregulated genes associated with chaperones and folding proteins observed in CB11 during beer fermentation, compared to the propagation step, at different times and the associated clustered biological processes from gene ontology analysis. Heatmap rows and columns were grouped using the Euclidean distance method and complete linkage.

*Subcellular localization of proteostasis- and chaperone-associated DEGs products*

The subcellular localization of proteostasis- and chaperone-associated DEGs products indicated that most of proteins can be found in cytoplasm, nucleus, ER, and mitochondria (Figures S7A to B and Figures S8A and C). In this sense, both DNA microarray analysis point to the same subcellular localization (Figures S8A and C) of different chaperone families whose members are upregulated in beer fermentation (Figures 8B and D).

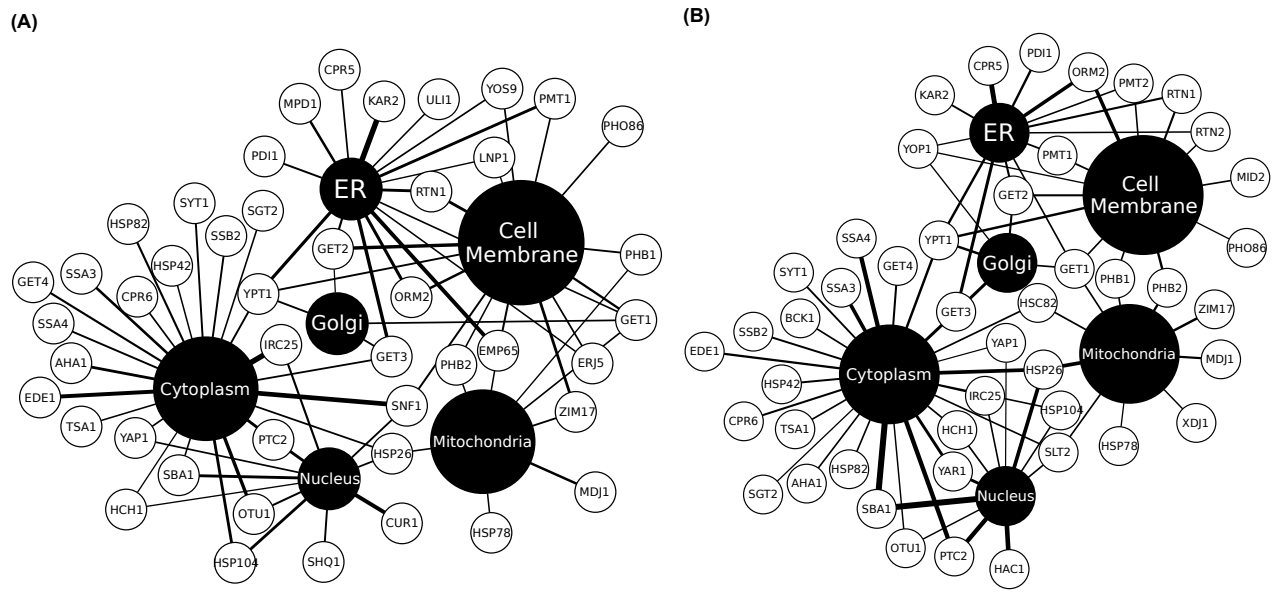

**Figure S7.** Networks describing the subcellular localization of proteostasis-coding DEGs obtained from DNA microarray single (GSE9423; A) and meta-analysis (GSE10205 versus GSE16376; B). The width of edges (thin to thick) is proportional to the mean logFC for each DEG evaluated in each analysis. The diameter of nodes representing the subcellular targets do not have any biological and/or statistical significance.

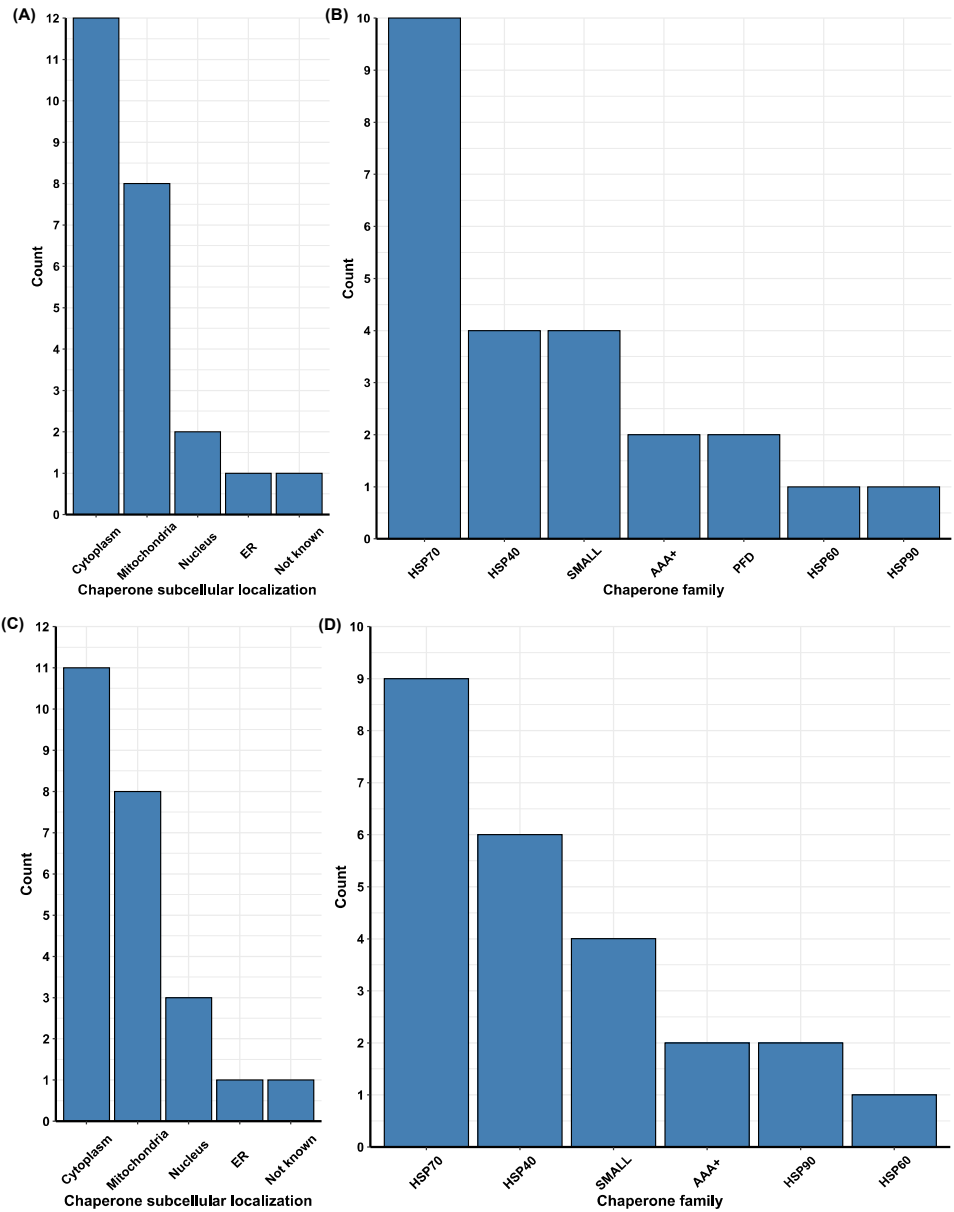

**Figure S8.** (A) and (C) Number of chaperones and folding protein coding genes found to be upregulated in different organelles of the lager yeast CB11 strain during beer fermentation, in comparison to yeast propagation as observed from DNA microarray single (GSE9423) and meta-analysis (GSE10205 versus GSE16376), respectively. (B) and (D) Number of coding genes upregulated in CB11 during beer fermentation, in comparison to yeast propagation, that are linked

to the major chaperone protein families as observed from DNA microarray single (GSE9423) and meta-analysis (GSE10205 versus GSE16376), respectively.

*Evaluation of proteostasis- and chaperone-associated Pan-DEGs*

In order to identify common upregulated proteostasis- and chaperone-associated DEGs (Pan-DEGs) in yeast lager CB11 strain during beer fermentation, DNA microarray data from single and meta-analysis were used (Figure S1). A high degree of overlap between DNA microarray analysis was observed, with 36 proteostasis-associated Pan-DEGs and 54 chaperone-associated Pan-DEGs identified (Figures S9A and B). This proteostasis- and chaperone-associated Pan-DEGs were further applied for biological processes and subcellular localization analyses (Figures 1 to 3 in the main text of the manuscript).

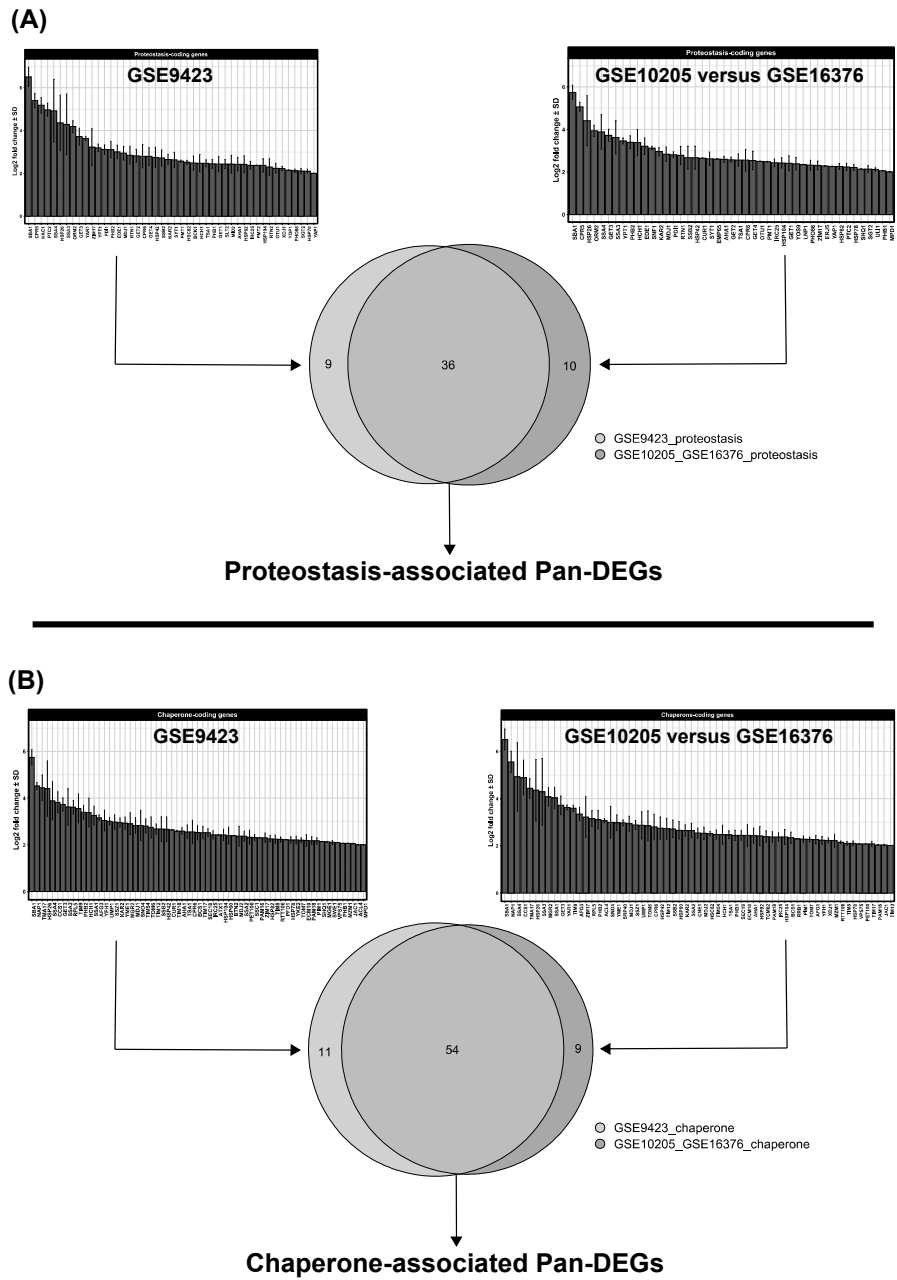

**Figure S9.** Evaluation of overlap degree of upregulated DEGs (Pan-DEGs) in yeast lager CB11 during beer fermentation between DNA microarray single- (GSE9423) and meta-analysis (GSE10205 versus GSE16376). (A) and (B), proteostasis- and chaperone-associated Pan-DEGs, respectively.

**Additional data**

*Crp6p-Rpd3p-Pbp1p interaction network*

A protein-protein interaction network of Crp6p with Rpd3p and Pbp1p (Figure S10) was obtained from STRING 11.0 (https://string-db.org). The following parameters were used for network prospection: *Saccharomyces cerevisiae* as selected organism; active prediction methods: databases and experiments; no more than five interactions in the first network shell; medium confidence score (0.400); evidence as meaning of network edges.

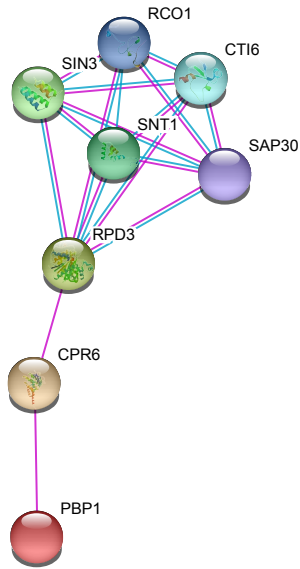

**Figure S10.** Protein-protein network interaction of yeast *Saccharomyces cerevisiae* Cpr6p with Rpd3p and Pbp1p. Edges color and number indicate supporting evidences from curated databases (light blue) and experiments (dark blue).

*Evaluation of fatty acid-associated DEGs in Saccharomyces pastorianus CB11 strain during beer fermentation*

Data from DNA single- (GSE9423) and meta-analysis (GSE10205 versus GSE16376) indicated that genes linked to fatty acid biosynthesis are upregulated in yeast lager CB11 strain during beer fernentation. In this sense, 124 upregulated fatty acid biosynthesis-associated DEGs were observed in GSE9423 dataset (Figure S11), while 113 DEGs were overexpressed in the GSE10205 versus GSE16376 datasets (Figure S12).

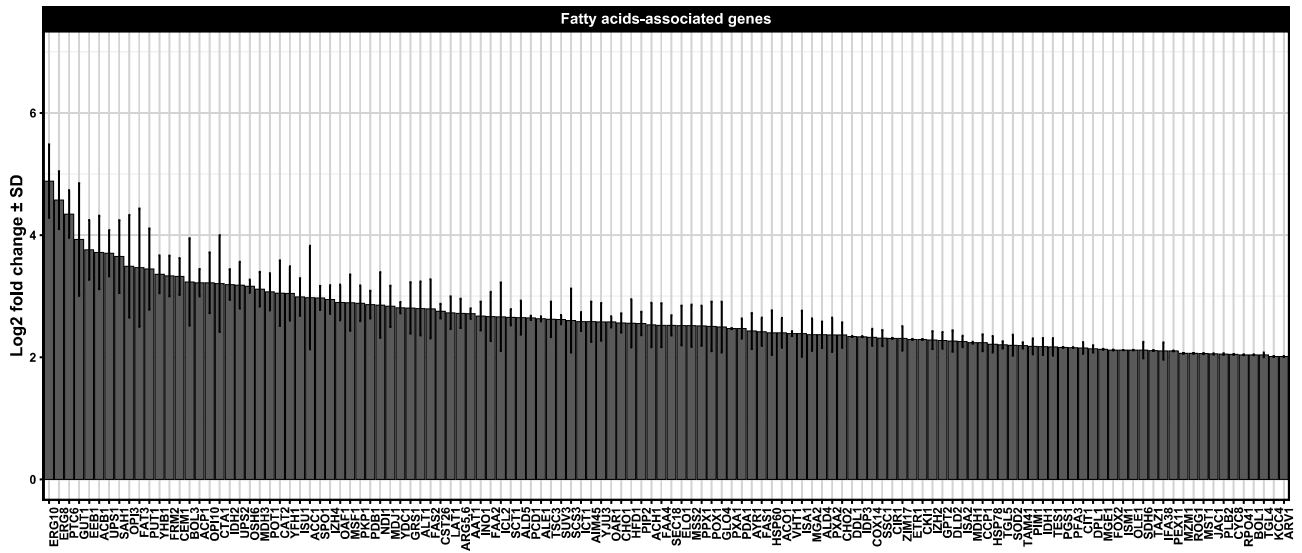

**Figure S11.** (A) Differentially upregulated genes from DNA microarray single analysis (GSE9423) associated with fatty acids biosynthesis observed in the lager yeast CB11 strain during beer fermentation, compared to the propagation step, at different times. The mean expression values are indicated by log2 fold change ± standard deviation (SD) on the y-axis and in the inset. Gene names are indicated on the x-axis.

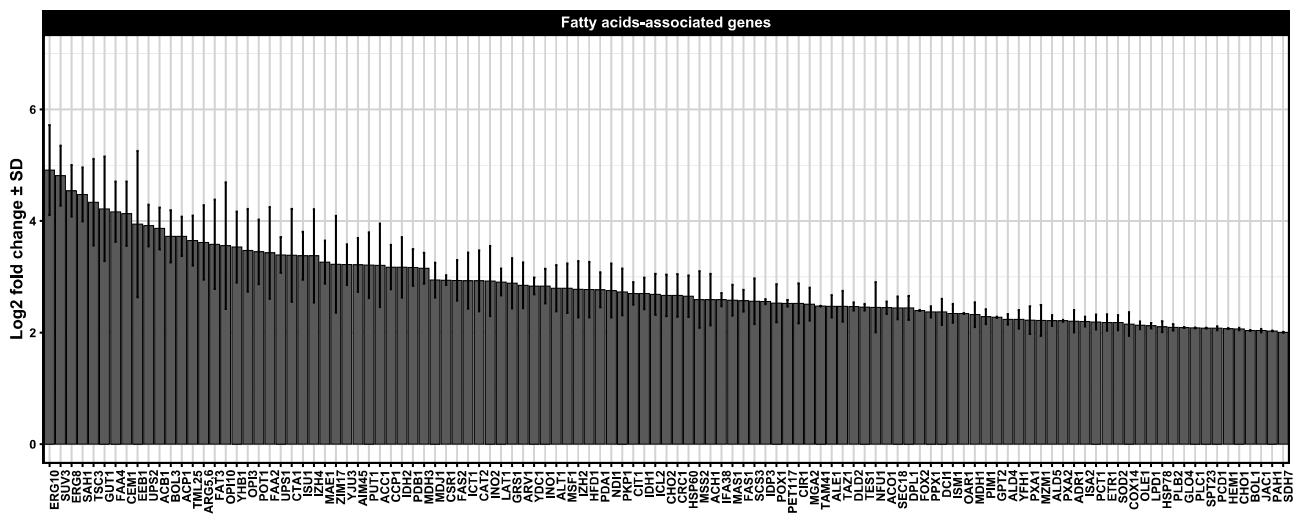

**Figure S12.** Differentially upregulated genes from DNA microarray meta-analysis (GSE10205 versus GSE16376) associated with fatty acids biosynthesis observed in the lager yeast CB11 strain during beer fermentation, compared to the propagation step, at different times. The mean expression values are indicated by log2 fold change ± standard deviation (SD) on the y-axis and in the inset. Gene names are indicated on the x-axis.